\documentclass[12pt,preprint]{aastex}

\slugcomment{ApJ, 2009, in press}

\shorttitle{The Spitzer Deep, Wide-Field Survey}
\shortauthors{Ashby et al.}

\newcommand{\SSS}{{\it Spitzer}}
\newcommand{\etal}{{et al.}~}

\newcommand{\Msun}{$M_\odot$}

\def\spose#1{\hbox to 0pt{#1\hss}}
\def\simlt{\mathrel{\spose{\lower 3pt\hbox{$\mathchar"218$}}
     \raise 2.0pt\hbox{$\mathchar"13C$}}}
\def\simgt{\mathrel{\spose{\lower 3pt\hbox{$\mathchar"218$}}
     \raise 2.0pt\hbox{$\mathchar"13E$}}}

\begin{document}

\title{The Spitzer Deep, Wide-Field Survey (SDWFS)}

\author
{
M.~L.~N.~Ashby\altaffilmark{1},
D.~Stern\altaffilmark{2},
M.~Brodwin\altaffilmark{1,3},
R.~Griffith\altaffilmark{2},
P.~Eisenhardt\altaffilmark{2},
S.~Koz{\l}owski\altaffilmark{4},
C.~S.~Kochanek\altaffilmark{4,5},
J.~J.~Bock\altaffilmark{6},
C.~Borys\altaffilmark{6},
K.~Brand\altaffilmark{7},
M.~J.~I.~Brown\altaffilmark{8},
R.~Cool\altaffilmark{9},
A.~Cooray\altaffilmark{10},
S.~Croft\altaffilmark{11},
A.~Dey\altaffilmark{12},
D.~Eisenstein\altaffilmark{13},
A.~H.~Gonzalez\altaffilmark{14},
V.~Gorjian\altaffilmark{2},
N.~A.~Grogin\altaffilmark{7},
R.~J.~Ivison\altaffilmark{15,16},
J.~Jacob\altaffilmark{2},
B.~T.~Jannuzi\altaffilmark{12},
A.~Mainzer\altaffilmark{2},
L.~A.~Moustakas\altaffilmark{2},
H.~J.~A.~R\"ottgering\altaffilmark{16},
N.~Seymour\altaffilmark{17},
H.~A.~Smith\altaffilmark{1},
S.~A.~Stanford\altaffilmark{18},
J.~R.~Stauffer\altaffilmark{19},
I.~Sullivan\altaffilmark{7},
W.~van Breugel\altaffilmark{20},
S.~P.~Willner\altaffilmark{1},
and 
E.~L.~Wright\altaffilmark{21} \\
}
\altaffiltext{1}{Harvard-Smithsonian Center for Astrophysics, 60 Garden St., Cambridge, MA 02138
[e-mail:  {\tt mashby@cfa.harvard.edu}]}
\altaffiltext{2}{Jet Propulsion Laboratory, California Institute of Technology, Pasadena, CA 91109}
\altaffiltext{3}{W. M. Keck Postdoctoral Fellow at the Harvard-Smithsonian Center for Astrophysics}
\altaffiltext{4}{Department of Astronomy, The Ohio State University, Columbus, OH 43210}
\altaffiltext{5}{The Center for Cosmology and Astroparticle Physics, The Ohio State University, Columbus, OH 43210}
\altaffiltext{6}{California Institute of Technology, Pasadena, CA 91125}
\altaffiltext{7}{Space Telescope Science Institute, Baltimore, MD 21218}
\altaffiltext{8}{School of Physics, Monash University, Clayton 3800, Victoria, Australia}
\altaffiltext{9}{Department of Astrophysical Sciences, Princeton University, Princeton, NJ 08544}
\altaffiltext{10}{University of California, Irvine, CA 92697}
\altaffiltext{11}{University of California, Berkeley, CA 94720}
\altaffiltext{12}{NOAO, 950 N. Cherry Ave., Tucson, AZ 85719}
\altaffiltext{13}{Steward Observatory, Tucson, AZ 85721}
\altaffiltext{14}{Department of Astronomy, University of Florida, Gainesville, FL 32611}
\altaffiltext{15}{UK Astronomy Technology Centre, Royal Observatory, Blackford Hill, Edinburgh, EH9 3HJ, UK}
\altaffiltext{16}{Institute for Astronomy, University of Edinburgh, Blackford Hill, Edinburgh, EH9 3HJ, UK}
\altaffiltext{17}{Leiden Observatory, Leiden University, PO Box 9513, 2300 RA Leiden, the Netherlands}
\altaffiltext{18}{Mullard Space Science Laboratory, University College London, 
Holmbury St. Mary, Dorking, Surrey RH5 6NT, UK}
\altaffiltext{19}{University of California, Davis, CA 95616}
\altaffiltext{20}{{\it Spitzer} Science Center, California Institute of Technology, Pasadena, CA 91125}
\altaffiltext{21}{University of California, Merced, CA 95344}
\altaffiltext{22}{University of California, Los Angeles, CA 90095-1562}

\begin{abstract}

The \SSS\ Deep, Wide-Field Survey (SDWFS) is a four-epoch infrared
survey of ten square degrees in the Bo\"otes field of the NOAO
Deep Wide-Field Survey using the IRAC instrument on the {\it Spitzer
Space Telescope}.  SDWFS, a \SSS\ Cycle~4 Legacy project, occupies
a unique position in the area-depth survey space defined by other
\SSS\ surveys.  The four epochs that make up SDWFS permit -- for
the first time -- the selection of infrared-variable and high proper
motion objects over a wide field on timescales of years.  Because
of its large survey volume, SDWFS is sensitive to galaxies out to
$z \sim 3$ with relatively little impact from cosmic variance for
all but the richest systems.  The SDWFS datasets will thus be
especially useful for characterizing galaxy evolution beyond $z
\sim 1.5$.  This paper explains the SDWFS observing strategy and
data processing, presents the SDWFS mosaics and source catalogs,
and discusses some early scientific findings.  The publicly-released,
full-depth catalogs contain 6.78, 5.23, 1.20, and $0.96\times10^5$
distinct sources detected to the average $5 \sigma$, 
4\arcsec\ diameter, aperture-corrected limits of 19.77, 18.83, 16.50, and
15.82 Vega mag at 3.6, 4.5, 5.8, and 8.0\,\micron, respectively.
The SDWFS number counts and color-color distribution are consistent
with other, earlier \SSS\ surveys.  At the 6~min integration time
of the SDWFS IRAC imaging, $>50\%$ of isolated FIRST radio sources
and $> 80\%$ of on-axis XBo\"otes sources are detected out to
8.0\,\micron.  Finally, we present the four highest proper motion
IRAC-selected sources identified from the multi-epoch
imaging, two of which are likely field brown dwarfs of mid-T
spectral class.

\end{abstract}


\section{Introduction}

In 2003, the IRAC Shallow Survey (ISS; Eisenhardt \etal 2004) mapped 8.5
deg$^2$ of the Bo\"otes field of the NOAO Deep Wide-Field Survey 
(hereafter NDWFS; Jannuzi \& Dey 1999).  
The ISS traced the evolution and clustering of massive galaxies 
(Brand \etal 2006; Brown \etal 2007, 2008; White \etal 2007), 
detected thousands of active galaxies (Stern \etal 2005; Desai \etal 2008)
and nearly 100 galaxy clusters at $z > 1$ (Stanford \etal 2005,
Brodwin \etal 2006, Elston \etal 2006; Eisenhardt \etal 2008), 
empirically extended low-resolution galaxy templates out to 
10\,\micron\ (Assef \etal 2008), defined the mid-infrared galaxy
luminosity functions for the local universe (Huang \etal 2008; 
Dai \etal 2009), and identified rare objects ranging 
from a radio-loud quasar at $z=6.12$ 
to a T4.5 field brown dwarf (Stern \etal 2007).  With a mere 90\,s
per pointing, this survey detected over 350,000 sources in the
mid-infrared (mid-IR) bands observed by IRAC.  These objects have a 
redshift distribution that spans most of the history of the 
universe (Brodwin \etal 2006).

Inspired by these successes, we undertook the \SSS\ Deep, Wide-Field
Survey (SDWFS; Figure~1), a 200\,hr Cycle~4 \SSS\ Legacy project.  SDWFS
re-imaged the same Bo\"otes field three more times during $2007 -
2008$ to the same depth at each epoch as the original IRAC Shallow Survey.  
The final SDWFS survey, doubling the photometric depth
of the original survey, occupies a unique position in area-depth
space relative to previous \SSS\ projects (Figure~\ref{fig.etendue})
and will be valuable for scientific investigations ranging from probing
the diffuse background from primordial galaxies to identifying the
coldest Galactic brown dwarfs.  By cadencing the IRAC observations,
SDWFS has also opened the largely unexplored territory of mid-IR
variability, a powerful tool for identifying and studying active
galaxies.  A program of this breadth has a wide range of astronomical uses. 
Below we highlight four of the key investigations that motivated 
the SDWFS project.  Future papers will describe each of these in 
more detail, but meanwhile we have made the dataset public and describe 
the potential herein.

{\it Diffuse Infrared Background from Population~III Stars.}  While
the first galaxies can not be individually detected with current
instruments, the ensemble of primordial galaxies produces a diffuse
extragalactic background, most prominently at near-infrared
wavelengths.  This background, which directly probes the earliest
phases of star formation in the Universe, has a distinct 
angular power spectrum that peaks at scales $\ell \approx 1500 \approx
7.5\arcmin$ (Cooray \etal 2004).  The background also has a distinct 
spectral energy distribution (SED), falling
sharply shortward of redshifted Ly$\alpha$ ($\lambda_{\rm obs}
\approx 1.3 - 2.6\,\mu$m).  Analyzing data from narrow ($\sim
10\arcmin \times 10\arcmin$), deep {\it Spitzer} fields, Kashlinsky
\etal (2005, 2007) reported detection of background
fluctuations from first-light galaxies at 3.6 to 4.5~$\mu$m.  These
exciting, yet controversial, results have profound cosmological
implications and have generated significant attention both in the
press and in the astronomical community.  In particular, there is
concern that a significant fraction of the reported background is
due to faint foreground sources, just below the point source
detection level (Cooray \etal 2007; Sullivan \etal 2007).  
The SDWFS dataset explores an ideal depth and area combination
for removing the signal from local galaxy correlations from
fluctuation maps (Bock \etal 2006).
A fundamental limitation of deep fields for studies of primordial
galaxy background fluctuations is that they are only two to three IRAC
pointings across, and therefore probe only $\ell \simgt 10^4$.  SDWFS,
with its wide-area, shallow coverage, is better-suited for studying
the diffuse background because it probes the first-galaxy fluctuation
spectrum on {\em both} sides of its predicted $\ell \approx 1500$ peak.
Deep observations are not essential; SDWFS is sensitive to
Population~III background fluctuation levels a factor of 20 fainter
than claimed by Kashlinsky \etal (2005), and even pessimistic
models suggest fluctuations must exist at these levels (Cooray \etal 2004).

{\it Galaxy Clustering at 0.5 $< z <$ 2.5.}  The growth of galaxies
is one of the most challenging problems in modern astronomy.  A red
sequence of early-type galaxies, with negligible rates of ongoing
star formation, is already in place by $z=1.4$ (e.g., Stanford \etal
2005).  Despite having little star formation, the stellar mass
within the red galaxy population increases by a factor of two or
more between $z\sim1$ and $z\sim0$, presumably due to the truncation
of star formation in blue galaxies (e.g., Bell \etal 2004; Brown
\etal 2007, Faber \etal 2007).  Where and why this occurs is unclear.
Cold dark matter models predict rapid growth of massive dark matter 
halos, even at late ($z < 1$) times.  However, observations show that 
the most massive galaxies grow slowly at late times compared to 
lower-mass galaxies (e.g., Bundy \etal 2006), a phenomenon commonly
referred to as ``downsizing.'' To
accurately measure how rapidly massive galaxies grow, one needs
large volumes to measure the space
density of galaxies with a precision of 10\% or better.  One can
then determine how galaxies are growing relative to their host dark
matter halos by measuring their clustering.  As the evolving space
density and clustering of dark matter halos are accurately modeled by
analytic approximations and simulations, one can use the observed
space density and clustering of galaxies to determine how galaxies
populate dark matter halos (e.g., Zehavi \etal 2005).  White \etal
(2007) used the original Bo\"otes field data to show that the
evolving clustering of galaxies is inconsistent with passive models
where massive galaxies do not merge over cosmic time.  Brown \etal
(2008) used the space density and clustering of galaxies to show
that, at late times, massive galaxies grow slowly relative to their host dark matter
halos, in part because much of the stellar mass within these halos
resides within satellite galaxies. The deeper SDWFS data will improve
the photometric redshifts for galaxies out to $z \sim 2.5$, allowing
the extension of these studies to the epoch where the global star
formation rate peaks. Furthermore, the deeper SDWFS data should
result in the detection of ten $z>1.5$ clusters with halo masses
of $10^{14}$\,\Msun\ or more.  These may be the environments in
which the first red galaxies formed.  Will these clusters contain
massive galaxies, or will we see evidence for the
cluster galaxies growing hierarchically from $z\sim 2$ until the
present day?

{\it IRAC Studies of AGN and AGN Variability.}  The primary goals
of the Bo\"otes AGN effort include studying the full (X-ray to
radio) SEDs of AGN (e.g., Hickox \etal 2007; Gorjian \etal 2008;
Higdon \etal 2008), to determine the infrared luminosity functions
of quasars (e.g., Brown \etal 2006), to study the physics of AGN
with less sensitivity to selection effects such as dust extinction
(e.g., Stern \etal 2005), to examine the distribution of AGN Eddington 
ratios (Kollmeier \etal 2006), and to probe the relation of AGN to their
large-scale environments (e.g., Galametz \etal 2009).   In particular,
AGN are known to exhibit photometric variability throughout the
electromagnetic spectrum.  By cadencing the IRAC observations over
multiple visibility windows, SDWFS has opened the largely unexplored
territory of mid-IR variability.  Dramatic AGN variations
have been seen at X-ray, optical, near-infrared, and radio wavelengths
(e.g., Hovatta \etal 2008; Papadakis \etal 2008; Sarajedini \etal
2006).  Monitoring campaigns are an efficient method to identify
AGN and provide information on the size scales, geometries,
and physics of the nuclear regions, providing fundamental tests of
unified AGN models.  The challenges of ground-based mid-IR
astronomy, however, have largely restricted infrared variability
studies to low-luminosity, nearby sources.  One study found
variability in most (39/41) sources, with variability most apparent
at longer wavelengths (Glass 2004).  The wide-area, shallow SDWFS
program is the ideal exploratory probe to identify the brightest,
most variable infrared sources for detailed study and multiwavelength
monitoring while mid-IR capabilities are still available.

{\it The Coldest Brown Dwarfs.}  One of the most important observational
topics in the field of sub-stellar objects is the detection and
characterization of objects cooler than T~dwarfs --- the so-called 
Y~dwarfs, cooler than about 700\,K.  Such objects must exist because
brown dwarfs with inferred masses down to $\approx 5$\,M$_{\rm Jup}$ have been
identified in star-forming regions.  
According to theoretical models,
dwarfs less massive than 30\,M$_{\rm Jup}$ with ages $> 4$\,Gyr should
have T$<$600\,K, but none has yet been found.  This is primarily
because their SEDs peak at $\approx 4.5\,\mu$m (e.g., Burrows \etal
2003), making them very faint at ground-based optical through
near-infrared wavelengths.  For instance, while a 600\,K brown dwarf
could be detected by 2MASS only if closer than $1$\,pc, it would
be detectable in a single SDWFS epoch out to almost 50\,pc at 4.5\,\micron.  
Moreover, the multi-epoch SDWFS survey 
can detect the proper motion of any dwarf with tangential
velocity $>$20\,km\,s$^{-1}$ out to 50\,pc.  The models of
Martin \etal (2001) predict that about 10\% of brown dwarfs should be
of type Y.  SDWFS ought to detect $\approx 5$ brown dwarfs cooler
than 1400\,K, and there is a reasonable chance of detecting a Y dwarf.
Our multi-epoch survey will thus provide a systematic census of faint
brown dwarfs.  We have already confirmed one T4.5 brown dwarf from
the original IRAC Shallow Survey (Stern \etal 2007), and we have
identified four additional mid-T brown dwarf candidates based on proper motion
measurements in the SDWFS data set (\S4.6).  A sample of
cooler brown dwarf candidates, later than approximately T7, has
recently been identified from the full SDWFS data set 
(Eisenhardt et al. 2009, in preparation).  Including the optical data, the Bo\"otes field's
13-year baseline will also provide a long lever arm for proper
motion studies of other (less red) Galactic populations.

\subsection{The Bo\"otes Field}

Much as the Great Observatory Origins Deep Survey (GOODS; Giavalisco
\etal 2004) fields have become the fields of choice for ultradeep
pencil-beam surveys across the electromagnetic spectrum, the 9
deg$^2$ Bo\"otes field has become the wide-area, deep survey field
of choice and thus was chosen as the target field for SDWFS.
Motivated by its low Galactic background and high ecliptic latitude,
various teams have mapped the entire field at X-ray wavelengths
with {\it Chandra} (Murray \etal 2005; Kenter \etal 2005; Brand
\etal 2006), at ultraviolet wavelengths with {\it GALEX} (Hoopes
\etal 2004), at optical/near-infrared wavelengths from NOAO (Jannuzi
\& Dey 1999; Elston \etal 2006), at mid-IR wavelengths with
{\it Spitzer} (Eisenhardt \etal 2004; Soifer \etal 2004), and at
radio wavelengths from the VLA and Westerbork at 20, 90, and 200~cm
(de~Vries \etal 2002; Croft \etal 2004, 2008).  Over 20,000 spectroscopic
redshifts in Bo\"otes have been obtained, the majority from the AGN
and Galaxy Evolution Survey (AGES; Kochanek et al. 2009, in preparation), and
the field is also the target of both ongoing and future surveys,
such as deeper near-infrared imaging with the NEWFIRM camera on the
Kitt Peak 4~m (led by co-I Gonzalez), 850~$\mu$m imaging with SCUBA2,
a far-infrared GTO survey with {\it Herschel}, near-infrared imaging
with the CIBER rocket experiment in Spring 2009 (led by co-I Bock),
and hard X-ray ($6 - 79$~keV) imaging with the {\it Nuclear
Spectroscopic Telescope Array} ({\it NuSTAR}), scheduled for launch
in 2011.  Most of these multiwavelength data are (or will be) public,
and over 80 refereed papers have been published from the Bo\"otes
surveys.\footnote{\tt 
http://www.noao.edu/noao/noaodeep/ndwfspublications.html.}

This paper is organized as follows:  \S2 discusses the details of
the SDWFS observing strategy and data reduction, and \S3 describes the
source identification, astrometric verification, and photometry.
Section 4 presents preliminary results from the analysis of the SDWFS
data, including mid-IR number counts, color distributions,
and identification fractions for both radio and X-ray sources.  
Section 5 summarizes these results.  Unless otherwise
stated, all magnitudes are in the Vega system.

\section{Observations and Data Reduction}

\subsection{Mapping Strategy}

The SDWFS observing campaign consisted of four co-extensive surveys
of the 8.5 square degree Bo\"otes field, carried out over roughly
four years (Table~1).  The first visit to the field occurred close
to the start of the \SSS\ mission, in 2004 January, as part of a
guaranteed time program led by the IRAC instrument team.  This
program was called the IRAC Shallow Survey (PID~30; Eisenhardt \etal
2004) and has had numerous successes, as discussed \S1.
A subsequent visit as part of the SDWFS program in 2007 August
remapped the same region to the same depth, and the final two visits,
again to the same depth, were at the beginning and end of the
following visibility window (2008 February and March).  Together, these four
observations provide temporal intervals of one month, six months,
and 3.5 years between visits (Table~\ref{tbl.epochs}).  
We refer to each of these four
Bo\"otes campaigns as an `epoch,' with the original IRAC Shallow
Survey now considered `epoch one' of SDWFS.

\begin{deluxetable}{clccc}
\tablecolumns{5}
\tablewidth{0pc}
\tablecaption{SDWFS Observations of the Bo\"otes Field.\label{tbl.epochs}}
\tablehead{
\colhead{Epoch} & \colhead{Observation Dates} & \colhead{PID} & \colhead{Pipeline Version} 
& \colhead{\# of BCDs}
}
\startdata
1 & 2004 Jan 10--14 &    30\tablenotemark{a} & 16.1.0 & 17014 \\
2 & 2007 Aug 8--13  & 40839 & 16.1.0 & 19956 \\
3 & 2008 Feb 2--6   & 40839 & 17.0.4 & 20200 \\
4 & 2008 Mar 6--10  & 40839 & 17.0.4 & 20680 \\
\enddata
\tablenotetext{a}{The IRAC Shallow Survey, Eisenhardt \etal (2004).}
\end{deluxetable}

The observing strategy was optimized to maximize the reliability
of the data taken within single epochs.  During each of the campaigns,
the entire Bo\"otes field was observed to a depth of 90\,s per sky
position.  The field was split into 15 `groups', each consisting of three 
coextensive Astronomical Observing Requests (AORs).  On average,
each group covered approximately 45\arcmin$\times$45\arcmin, or 
$9\times9$ IRAC pointings.  In practice, group size and shape varied
because of the gerrymandering needed to fit groups to the chevron shape 
of the existing optical/near-infrared coverage of the NDWFS.

To ensure that flux from asteroids could later be reliably masked and 
excluded from the SDWFS mosaics, each group was observed
in three passes of 30\,s depth, with each pass of each group
constituting a separate AOR.  The 
30\,s frame time provided background-limited or nearly background-limited
data in all four IRAC bands and ensured that 
dithering overheads did not dominate the AOR execution time.
The time required
to obtain a single 30\,s pass on a group ensured gaps of at least
2\,hr between observations of each sky position.  For typical asteroid
motions of 25\arcsec\, hr$^{-1}$, asteroids moved by $\approx
1\arcmin$ between maps.  Since this is much smaller than a typical
map width, this mapping strategy reliably tracks asteroid motions,
thus providing both a clean, final mosaic for Galactic and extragalactic
studies and a scientifically valuable asteroid sample at
high ecliptic latitude.

Our mapping strategy incorporated several elements to facilitate
self-calibration of the data by maximizing inter-pixel correlations
(e.g., Arendt \etal 2000).  We {\em dithered} the observations on
small scales and {\em offset} by one-third of an IRAC field of
view between successive passes through each group.  This provided
inter-pixel correlation information on both small and large scales.
In addition, for AOR groups having rectangular shapes, we {\em cadenced} the
observations such that revisits cover the same area but with a
different step size.  For example, a group that imaged a 
40\arcmin$\times$60\arcmin\, region using an $8 \times 12$ grid map with
290\arcsec\, step size subsequently used $9\times12$ and $8\times13$ 
grid maps (with suitably scaled down step sizes).  With
a $\simlt 10\%$ penalty in mapping efficiency, cadencing significantly
enhances the inter-pixel correlations across all scales and
greatly increases the redundancy of the datasets.  
Finally, by reobserving the field
multiple times at different roll angles and different times, our
observing strategy is very robust against bad rows and columns,
large-scale cosmetic defects on the array, glints, changes in the
zodiacal background, after-images resulting from saturation due to
bright stars, variations in the zero level, and the color dependence
of the IRAC flat-field across the array (Hora \etal 2008).  In
particular, the challenging diffuse background measurements, one
of the key goals of the SDWFS project (e.g., Sullivan \etal 2007),
are vastly aided by the redundant coverage:  independent data sets
are the best way to assess many systematic errors.  
The final SDWFS coverage is $3\times30$\,s for each epoch and 
$12\times 30$\,s for the complete survey.  The number of exposures
needed to cover the field in a single epoch varied from 17,014 to 20,680, depending
on roll angle.  The full, co-added SDWFS 3.6\,$\mu$m map is presented
in Figure~\ref{ch1_image}, and the corresponding coverage map is presented in
Figure~\ref{cvg_map}.  Figure~\ref{fig:cutouts} illustrates how the 
source density changes through the IRAC bands at the full SDWFS depths. 
Figure~\ref{fig:coverage} presents the cumulative area 
coverage as a function of exposure time.  

The data reduction was based on the IRAC Basic Calibrated Data 
(BCD).  The 3.6, 5.8, and 8.0\,micron\ BCD frames were object-masked 
and median-stacked on a per-AOR basis; the resulting stacked 
images were visually inspected and subtracted from individual 
BCDs within each AOR to eliminate long-term residual images 
arising from prior observations of bright sources.
This was not necessary for the 4.5\,\micron\ BCDs because the 4.5\,\micron\ IRAC detector
array does not suffer from residual images.

The first two epochs were processed with pipeline version 16.0.1.
The 3.6 and 4.5\,\micron\ BCDs were examined by hand and modified
using custom software routines to correct column pulldown and
multiplexer bleed effects associated with bright sources.  Much of
this software is available from the \SSS\ Science Center\footnote{\tt
http://ssc.spitzer.caltech.edu/archanaly/contributed/browse.html}
and is based on algorithms designed and coded by SDWFS co-I's.  The
last two epochs were processed with pipeline version 17.0.4.  For
these, we used the so-called corrected BCD frames, in which the BCD
pipeline automatically applies multiplexer bleed and column pulldown
corrections.  The photometric calibrations are identical for both 
pipelines.

After these preliminaries, the data were mosaiced into eight
overlapping sub-fields (tiles), each slightly larger than a degree
across, using IRACproc (Schuster \etal\ 2006).  In order to take
some advantage of the subpixel shifts of our mapping strategy, the
mosaics were resampled to 0\farcs84 per pixel, so that each mosaic
pixel subtends half the area of the native IRAC pixel.  IRACproc
augments the capabilities of the standard IRAC reduction software
(MOPEX).  The software was configured to automatically flag and
reject cosmic ray hits based on pipeline-generated masks together
with a sigma-clipping algorithm for spatially coincident pixels.
IRACproc calculates the spatial derivative of each image and adjusts
the clipping algorithm accordingly.  Thus, pixels where the derivative
is low (in the field) are clipped more aggressively than are pixels
where the spatial derivative is high (point sources).  This avoids
downward biasing of point source fluxes in the output mosaics.
For the final, total coadd ($12\times30$\,s) of all four epochs, we proceeded in the same
manner as for the individual epochs with the minor change that only
temporal outlier rejection was performed.  

For both the single-epoch and the total coadds, the outlier rejection 
threshold was iteratively refined by inspecting difference images 
(\S\ref{ssec:irvar}): we subtracted successive epochs from 
each other to suppress emission from the non-varying celestial 
sources (i.e., almost all of them), so that cosmic ray artifacts 
became obvious against the dramatically flattened background.  

Each tile was reduced independently of all other tiles.  IRACproc
automatically identified and incorporated all exposures overlapping the 
specified field of a given tile into the corresponding mosaic.  
This inevitably resulted in each tile having some amount of irregular, 
low-coverage `crust' on its periphery even though the depth of
coverage within each tile's analysis region was complete and free of
cosmic rays by virtue of the iterated outlier rejection described
above.  The tiles were subsequently combined 
into full-field images, one for each
band and each epoch, using the Montage toolkit 
(ver.~3.0; Berriman \etal 2004).  The
coverage maps were used as a means of
preventing the low-coverage peripheral `crusts' 
--- with their relatively high prevalence of cosmic ray hits --- from
degrading the final mosaics.  The outcome was a suite of 20 
mosaics: four from each of the four epochs, plus another four for
the total 12$\times$30\,s coadd.  These data, including coverage
maps and catalogs, are all available from the \SSS\ Science Center.\footnote{\tt
http://ssc.spitzer.caltech.edu/legacy/sdwfshistory.html}

\section{Source Identification}
\subsection{Source Extraction and Photometry}

The survey depths achieved are well above the confusion limits even
when all four epochs are combined into total coadds comprising
12$\times$30\,s coverage.  For this reason we constructed source
catalogs using SExtractor (ver. 2.5.0; Bertin \& Arnouts 1996),
a standard tool for these purposes.
In addition to source catalogs, we also generated both
background and object `check-images.' Inspecting those check-images,
we iteratively chose parameter settings, ultimately settling on the
values listed in Table~\ref{tab:settings}.  The outcomes did not 
depend sensitively on many of the parameters, although the correct
choices of convolution filter width and deblending parameters were
necessary in order to obtain `object' check images that appeared
complete upon inspection.  We used the
coverage maps as detection weight images in the standard way.
Regions covered by fewer than two exposures were excluded using
separate `flag' images constructed from the coverage maps.

\begin{deluxetable}{lcccc}
\tablecaption{SExtractor Parameter Settings.\label{tab:settings}}
\tablewidth{0pt}
\tablehead{
\colhead{} & \colhead{3.6\,\micron} & \colhead{4.5\,\micron}& \colhead{5.8\,\micron} & \colhead{8.0\,\micron}}
\startdata
DETECT\_MINAREA [pixel] & 3 & 3 & 4 & 4 \\
FILTER & gauss\_2.0\_5x5 & gauss\_2.0\_5x5 & gauss\_2.0\_5x5 & gauss\_2.5\_5x5 \\
DEBLEND\_MINCONT & 0.00005 & 0.00005  & 0.0001 & 0.0001\\  
SEEING\_FWHM [arcsec] & 1.66 & 1.72 & 1.88 & 1.98 \\ 
\enddata
\tablecomments{Parameter settings used identically in all four IRAC bands
were DETECT\_THRESH=1.5, DEBLEND\_NTHRESH=64, BACK\_SIZE=256,
BACK\_FILTERSIZE=3, BACKPHOTO\_TYPE=GLOBAL, and WEIGHT\_TYPE=MAP\_WEIGHT.}
\end{deluxetable}

Photometry was carried out using several approaches available in
SExtractor.  We compiled fluxes and magnitudes within ten apertures
having diameters from 1 to 10\arcsec\ in 1\arcsec\ increments, plus
15 and 24\arcsec, for twelve apertures in all.  We also measured
MAG\_AUTO magnitudes, for which SExtractor measures sources' fluxes
interior to elliptical apertures with sizes and orientations
determined using the second-order moments of the light distribution
measured above the isophotal threshold.

Having first verified that the astrometry is accurate on scales
much smaller than a SDWFS mosaic pixel (\S~\ref{sec:astrometry}),
we used SExtractor in dual-image mode to photometer the
SDWFS mosaics.  In dual-image mode, sources are detected, their
centers are located, and their apertures are defined in one image,
and subsequently photometry is carried out in another image using
those pre-established apertures and source centroids.   We used the
four-epoch 12$\times$30\,s mosaics as the detection images in all cases 
because of their superior noise properties.

SExtractor was configured to define a source as a set of three or more 
connected pixels each lying 1.5$\sigma$ above the background.  With this 
parameter setting, we detected a total of 8.2, 6.7, 3.1, and $1.8\times10^5$ sources 
at 3.6, 4.5, 5.8, and 8.0\,\micron, respectively, in the four-epoch
12$\times$30\,s SDWFS mosaics.  This corresponds to 1$\sigma$ depths of
21.7, 20.8, 18.7, and 18.0 Vega mag.

\subsection{SDWFS Catalogs}

To compensate for flux falling outside the SExtractor
apertures, we used the twelve-aperture photometry for
bright, unresolved sources in the full-depth, four-epoch
mosaics to derive empirical aperture corrections that correct
the photometry to 24\arcsec-diameter apertures.
These aperture corrections are listed in Table~\ref{tbl.apcorr}.
Comparing our 5\arcsec\ to 24\arcsec\ diameter aperture corrections
to the 4\farcs9 to 24\farcs4 diameter aperture corrections provided
in the IRAC Instrument Handbook (ver.3.0), our corrections are
systematically $\sim$3.5\% larger.  Besides the slight
differences in aperture radii, this difference likely arises from
a combination of the slightly different point spread function (PSF)
of our multi-epoch stacked images and differences in how the
backgrounds were determined.  In particular, stacking multi-epoch
data will suppress azimuthal asymmetries such as diffraction spikes,
leading to a PSF profile that is more closely circular.

\begin{deluxetable}{clrcc}
\tablecaption{SDWFS Aperture Corrections.\label{tbl.apcorr}}
\tablewidth{0pt}
\tablehead{
\colhead{Aperture Diameter} & \colhead{3.6\,\micron} & \colhead{4.5\,\micron} & \colhead{5.8\,\micron}   & \colhead{8.0\,\micron}
}
\startdata
$3\arcsec$ & $-$0.67 & $-$0.69 & $-$0.89 & $-$1.01 \\
$4\arcsec$ & $-$0.38 & $-$0.40 & $-$0.57 & $-$0.69 \\
$5\arcsec$ & $-$0.24 & $-$0.25 & $-$0.38 & $-$0.51 \\
$6\arcsec$ & $-$0.17 & $-$0.17 & $-$0.26 & $-$0.38 \\
\enddata
\tablecomments{Corrections (magnitudes) derived from well-detected 
stars in the full-depth, stacked SDWFS image.  The corrections have been 
added to the corresponding aperture photometry measurements in the SDWFS
catalogs to correct them to 24\arcsec\ diameter apertures, as used for
calibration stars.}
\end{deluxetable}

We merged the single-band source lists by combining the astrometry 
defined in the `detection' images together with photometry carried out 
in the `measurement' images.  
In this way we created a final set of 20 band-merged catalogs, i.e.,
for sources selected in each of the four IRAC filters, we merged
the four-band photometry from each of the four
epochs as well as from the total coadd.
Each catalog contains aperture-corrected photometry
measured within sets of identical circular apertures having diameters
of 4\arcsec\ and 6\arcsec plus MAG\_AUTO.  

Depending on the application, it may be advantageous to perform source
selection in different bands.  For example, selection at 3.6, 4.5, and 
8.0\,\micron\ might well be optimal for studies of nearby galaxies, for 
selection of substellar objects, and selection of red, high-redshift
objects, respectively.  Rather than impose our own selection bias on
users of SDWFS data, we make all 20 SDWFS band-matched catalogs available 
in the electronic edition of this article.  The catalog formats are defined
in Tables~\ref{tbl:format1}--~\ref{tbl:format20}.

\subsection{Completeness and Depth Simulations}
\label{sec:completeness}

The survey completeness in each band was assessed with a standard
Monte Carlo approach.  Using the IRAF task {\em noao.artdata.mkobjects},
ten thousand artificial stars in each half magnitude bin between
$10 \le \mbox{m(IRAC)} \le 22.5$ were added at random positions to
the mosaics.  Catalogs were extracted from the resulting 25 images
in each band exactly as described above.  The recovery rates as a
function of magnitude in the IRAC bands are plotted in
Figure\,\ref{brodwin_sims}.  
The recovered fraction has a sharp fall-off in the relatively sparse,
redder bands, while in the deeper 3.6 and 4.5\,$\mu$m data, the decline is more
gradual.  This is likely caused by moderate levels of confusion in
these more sensitive bands, where the high object density leads to
difficulties for object recovery in crowded regions.
The 50\%  and 80\% completeness levels are given in Table~\ref{tbl.completeness}.

\begin{deluxetable}{clrcc}
\tablecaption{SDWFS Completeness Limits.\label{tbl.completeness}}
\tablewidth{0pt}
\tablehead{
\colhead{Completeness Level} & \colhead{3.6\,\micron} & \colhead{4.5\,\micron} & \colhead{5.8\,\micron}   & \colhead{8.0\,\micron}
}
\startdata
80\% & 18.2 & 18.1 & 16.8 & 16.1 \\
50\% & 19.9 & 19.1 & 17.2 & 16.5 \\
\enddata
\tablecomments{Magnitudes are relative to Vega, and correspond to 
completeness levels established with the simulations described in
Section~\ref{sec:completeness}.
}
\end{deluxetable}


The mosaics' depths were also estimated using Monte Carlo techniques.
In all four IRAC bands, ten thousand random positions within the
mosaics (both single-epoch and SDWFS total) were photometered in
various apertures.  Gaussians were fitted to the negative half of
the measured surface brightness distributions, which are devoid of
any contribution from real sources, in order to reliably estimate
sky noise.  Table~\ref{tbl.depths} lists the aperture-corrected
5$\sigma$ depths in several apertures for both the first epoch and
the full survey.  The single-epoch depths can be a few tenths 
of a magnitude better than those reported from
the original IRAC Shallow Survey (Eisenhardt \etal 2004), primarily
at 3.6\,\micron.  This is attributable to improved processing algorithms.
In a corrected 4\arcsec\
diameter aperture from a single epoch, SDWFS reaches depths of 19.1, 18.2,
15.8, and 15.1\,mag in the IRAC bands.  The full-depth images reach
approximately 0.7~mag deeper.  

\begin{deluxetable}{clrcc}
\tablecaption{5$\sigma$ SDWFS Depths.\label{tbl.depths}}
\tablewidth{0pt}
\tablehead{
\colhead{Aperture Diameter} & \colhead{3.6\,\micron} & \colhead{4.5\,\micron} & \colhead{5.8\,\micron}   & \colhead{8.0\,\micron}
}
\startdata
\cutinhead{Single-Epoch}
$3\arcsec$ & 19.31 & 18.27 & 15.88 & 15.18 \\
$4\arcsec$ & 19.12 & 18.18 & 15.78 & 15.12 \\
$5\arcsec$ & 18.91 & 18.02 & 15.65 & 15.00 \\
$6\arcsec$ & 18.73 & 17.81 & 15.52 & 14.89 \\
\cutinhead{Total SDWFS}
$3\arcsec$ & 19.95 & 18.98 & 16.66 & 15.85 \\
$4\arcsec$ & 19.77 & 18.83 & 16.50 & 15.82 \\
$5\arcsec$ & 19.56 & 18.66 & 16.37 & 15.68 \\
$6\arcsec$ & 19.32 & 18.49 & 16.29 & 15.57 \\
\enddata
\tablecomments{Magnitudes are relative to Vega.  Depths, estimated
from the sky noise, are aperture-corrected and apply to survey locations
having the median exposure times of 90 and 420\,s in the single-epoch
and full-depth SDWFS mosaics, respectively.}
\end{deluxetable}

Single-epoch survey completeness (Fig.~\ref{griffith1}) 
was estimated by counting the epoch-four detections 
having 4\arcsec\ diameter aperture magnitudes that agreed within 0.2\,mag with
the corresponding detections in the full-depth SDWFS mosaics.  
The four 3$\times$30\,s SDWFS catalogs are $\sim$80\% complete at
the corresponding 5$\sigma$ sensitivity limits.
Single-epoch catalog unreliability (Fig.~\ref{griffith2}) 
was estimated by computing the percentage of sources in the 
epoch-four catalog which do {\sl not} appear in the full-depth mosaics.
At 3.6 and 4.5\,\micron, the single-epoch SDWFS catalogs are at least 99.7 and
99.9\% reliable at or above the 5$\sigma$ sensitivity limits.  At 5.8 and
8.0\,\micron\ the catalogs are 99.4\% reliable at these levels.

\subsection{Astrometry}
\label{sec:astrometry}

To estimate the accuracy of the astrometry in our SExtractor catalogs,
we matched sources selected at 3.6\,\micron\ (where IRAC has the highest spatial
resolution) to the 2MASS and USNOB\,1.0 (Monet \etal 2003) catalogs.
Figures~\ref{astrometry.2mass} and \ref{astrometry.usnob} show the results.

Within the 2\farcs0 matching radius, we matched all 16,045 Bo\"otes 
2MASS sources to our IRAC 3.6\,\micron\ source list.  Given that the 
SDWFS data are significantly deeper than 2MASS, the 100\% match rate 
was expected.  The astrometric agreement is excellent: the mean offsets 
(2MASS - SDWFS) are $0\farcs04$ and $-0$\farcs02 in Right Ascension and 
Declination, respectively.  The corresponding standard deviations of 
the measured offsets were 0\farcs3 in both coordinates, consistent with 
the expected measurement error given the IRAC spatial resolution of 
1\farcs66 (FWHM) and the 0\farcs84 mosaic pixel size.  
Given that the {\it Spitzer}/IRAC pipeline automatically assigns a 
coordinate solution to individual BCD frames by identifying and referencing 
the ensemble of all 2MASS stars imaged simultaneously in both IRAC FOVs
(the so-called super-boresight calibration), this level of accuracy 
is unsurprising.

An additional, independent estimate of SDWFS astrometric accuracy comes
from comparison with USNOB1.0.
There appears to a small but systematic coordinate offset between SDWFS
sources and those tabulated in USNOB1.0.  Median offsets (USNOB $-$
SDWFS) are 0\farcs06 and 0\farcs08 in right ascension and declination,
respectively, as shown in Figure~\ref{astrometry.usnob}.  The dispersion of offsets is
larger than for 2MASS,  0\farcs5 standard deviation in each
coordinate.  Thus the astrometric agreement of all three catalogs is
excellent overall, but there is a slight offset between 2MASS and SDWFS
relative to USNOB1.0 in this region ofsky.  The magnitude of the
discrepancy is somewhat smaller than that found between, e.g.,
zBo\"otes and NDWFS (0\farcs2; Figure 4 of Cool 2006).

\section{Analysis}
\subsection{Confusion}
\label{ssec:confusion}

We estimated worst-case confusion arising from {\sl well-detected} 
sources as follows.  SDWFS detected 6.78, 5.23, 1.20, and
$0.96\times10^5$ sources above the aperture-corrected 4\arcsec\ diameter
$5 \sigma$ threshholds (Table~5) in the full-depth mosaics covering
10.6\,deg$^2$.  The source density is greatest at 3.6\,\micron, 
roughly 0.008 per 3.6\,\micron\ beam (where the beam solid angle $\Omega$ 
was estimated under the assumption that in the four-epoch coadd, the trefoil 
IRAC PSF has a Gaussian profile with FWHM$=1\farcs66$, 
so that $\Omega=\pi \sigma^2 = 1\farcs57$\,arcsec$^2$).
This implies that fewer than 1\% of the 3.6\,\micron\ SDWFS sources are 
confused with other sources that lie above the 3.6\,\micron\ detection limit.

We also estimated source confusion arising from {\sl faint} sources
by examining the deep 3.6\,\micron\ source counts in the IRAC
image of the Extended Groth Strip (Barmby \etal 2008).  
Specifically, we summed EGS counts in the three half-magnitude 
bins lying just below the SDWFS 5$\sigma$ detection limit 
to arrive at a density of roughly 
$4.2\times10^5$ sources\,deg$^{-2}$ down to 21.7\,mag,
which corresponds roughly to the $1\sigma$ SDWFS limit.
This implies that if one extends the counts down to 
the $1\sigma$ level, there are about 0.05 sources per 
IRAC beam in the four-epoch 3.6\,\micron\ SDWFS mosaic. 
Of course, this estimate slightly underestimates the effects
of confusion, because galaxies are clustered rather than
randomly distributed, and they are not, in general, point sources.
Nonetheless, to a reasonable approximation roughly one out of 
every 20 SDWFS sources will have 3.6\,\micron\ photometry unavoidably 
contaminated by emission from a faint, overlapping source.  The percentage
will be smaller at the longer wavelengths.  This contamination is 
of course more significant for fainter sources, and is relatively 
unimportant for the high-significance detections.  

\subsection{Number Counts}

A fundamental feature of any survey is the relationship it reveals between 
the fluxes of the objects it detects and their area density, i.e., 
the source counts.  Accordingly, we plot the differential SDWFS number counts 
in Figure~\ref{fig:sdwfs_counts}.  The panels illustrate 
total counts (stars plus galaxies) in all four IRAC bands.  
For comparison, the stellar counts from Fazio \etal (2004) are also shown.  
Stars dominate the counts at the bright end of the distribution, 
but at the faint end, the counts are dominated by galaxies.

Figure~\ref{fig:sdwfs_counts} also compares the SDWFS count
distribution to that of Fazio \etal (2004).  The latter are drawn
from the original IRAC Shallow Survey and constitute the first SDWFS
epoch.  The agreement is excellent throughout, although the deeper
SDWFS data permit the counts to be estimated to fainter levels
than was originally possible for this field.  Both sets of counts
exhibit a change in slope at $[3.6]\approx17$, consistent with what
is seen in other work (e.g., Sullivan \etal 2007).  The point of
inflection seen in the 3.6\,\micron\ counts is consistent with the
models of Gardner (1998), which extrapolate $K$-band counts into
the IRAC 3.6\,\micron\ band.  Essentially, the negative K-correction
of a passively evolving population formed at high redshift ($z
\simgt 3$) produces a roughly constant apparent magnitude in the
shorter wavelength IRAC bands for $0.7 \simlt z \simlt 2$ (see
Fig.~1 of Eisenhardt \etal 2008).

At the faint end, the SDWFS counts are in excellent agreement with those
drawn from the much deeper IRAC survey of the Extended Groth Strip (Barmby \etal 2008).  
The differences appear when the counts approach the $5\sigma$ limit, where
the SDWFS incompleteness corrections become large and less certain.

\subsection{IRAC Color Distribution}

For the galaxy populations that are detected in wide-area, relatively 
shallow surveys like SDWFS, a limited number of emission processes 
dominate the IRAC SEDs.  These include stellar thermal emission from 
bulge and/or disk populations, to which the 3.6 and 4.5\,micron\ bands 
are most sensitive.  There is also typically a contribution of 
7.7\,\micron\ PAH emission, mostly from low-redshift galaxy disks, 
as well as the predominantly red emission from AGN that often takes 
the form of a power law.  These contributions to the IRAC SEDs are 
reflected in the way SDWFS sources are distributed in the IRAC 
color-color space shown in Figure~\ref{fig:colors}.  Broadly speaking, 
their distribution is consistent with that shown in other surveys.
A number of sources fall near the origin in this color-color space;
these are the stars and low-redshift galaxies dominated by starlight (e.g., elliptical
galaxies).  Indeed, the colors of elliptical galaxy NGC\,4429 intersect 
this locus of points at modest redshift.  Sources that are red in [5.8]$-$[8.0] 
and relatively blue at shorter wavelengths ([3.6]$-$[4.5]$<0.5$) 
are consistent with emission from nearby galaxies rapidly forming stars.  Objects like these 
generate significant
7.7\,\micron\ PAH emission and therefore appear red to IRAC.  The
$0\le z \le 2$ non-evolutionary track shown for M\,82 in this color-color 
space follows the distribution of these sources remarkably well.

Finally, the portion of Figure~\ref{fig:colors} enclosed within
the dotted line (the `Stern wedge'; Stern \etal 2005; 
see also Lacy \etal 2004) is a regime
that by construction contains galaxies that are red in both [5.8]$-$[8.0] 
and [3.6]$-$[4.5].  At redshifts below 1.5, objects that lie in 
this region of color-color space have mid-IR SEDs that are 
closely approximated by a power law and are predominantly AGN.
A large fraction of the X-ray-detected sources in Bo\"otes lie
in this region (Gorjian \etal 2008).
While it is true that high-redshift ($z>2$) galaxies can enter
the Stern \etal AGN region (e.g., Stern \etal 2005; Barmby \etal 2008), such objects
generally have 5.8 and 8.0\,\micron\ fluxes that fall below 
the SDWFS flux limits, so they are thought not to be a significant
contaminant at SDWFS depths.

\subsection{Mid-Infrared Detections of Radio Sources}

The FIRST (Faint Images of the Radio Sky at Twenty-Centimeters;
Becker, White, \& Helfand 1995) survey mapped 9055 square degrees
of the North and South Galactic caps with the Very Large Array,
reaching a typical $5 \sigma$ detection threshold of 0.75\,mJy at
1.4\,GHz.  This large survey, which includes the Bo\"otes field, has
a source density of $\sim 90$ sources per square degree and a typical
positional accuracy of 1\arcsec.  Approximately 30\% of the sources
have optical identifications in the Sloan Digital Sky Survey 
(Ivezi{\'c} \etal 2002).  Here
we consider the IRAC properties of FIRST sources as observed by SDWFS.

According to the latest catalog (08oct31), 887 FIRST sources fall
in the 3.6\,$\mu$m SDWFS coverage map.  While several techniques
have been presented in the literature for cross-correlating radio
catalogs with catalogs at shorter wavelength, Sullivan \etal (2004)
demonstrated that positional coincidence is sufficient when 
both the radio and shorter wavelength catalogs have high positional
accuracies.  We therefore adopted a simple, empirically-derived
2\arcsec\ match radius to identify IRAC counterparts to FIRST
sources.  An identical match radius was used by El Bouchefry \&
Cress (2006) when matching FIRST to the NDWFS optical and near-IR
Bo\"otes data.  Given the source density of $5 \sigma$ IRAC sources from the
stacked SDWFS data set, this match radius will yield a 7.3\% spurious
identification rate at the $5\sigma$ limit of the 3.6\,$\mu$m images (4\arcsec\ diameter
apertures), falling to 1.0\% at 8.0\,$\mu$m.
The spurious match rate will, of course, be smaller for brighter sources.
We find 3.6\,$\mu$m identifications for 570 (64.3\%) of the FIRST
sources.  However, approximately 30\% of FIRST sources have resolved
structure on scales of $2 - 30\arcsec$, with the classic example
being a double-lobed radio source comprised of two radio-emitting
lobes separated by several arcseconds straddling a host
galaxy.  Our simple position matching scheme will yield no (correct)
identifications for such a configuration.  We have attempted to
minimize the number of such systems by considering only isolated
FIRST sources, defined to be at least 30\arcsec\ from the next
nearest FIRST source.  The average FIRST source density implies
only a 2\% chance that a FIRST source will have a random, unrelated
source within that radius.  Of the 887 FIRST sources in SDWFS, 667
are thus defined as isolated, of which 512 (76.8\%) have IRAC
3.6\,$\mu$m identifications.  Figure\,\ref{fig:radioIDs} shows how
this identification rate depends both on IRAC passband and mid-IR
survey depth for the isolated FIRST sources.  Compared
to the original IRAC Shallow Survey, the deeper SDWFS data provide
a modest improvement for the 3.6 and 4.5\,$\mu$m identification
fractions but approximately double the number of 5.8 and 8.0\,$\mu$m
identifications.  The resulting identification fraction ($>50$\%) 
allows us to probe the IRAC color-color
distribution of radio sources with higher fidelity
(Figure\,\ref{fig:xrayradio}).  
El Bouchefry \& Cress (2006) matched five square degrees of 
the Bo\"otes NDWFS field with the FIRST catalog and found
near-infrared identifications for only 40\% of FIRST sources 
(to $K \sim 19.4$) and optical identifications for $\sim 72\%$ of FIRST
sources (to $B_{\rm W},R,I \sim 25.5$).
Approximately 30\%\ of the FIRST sources with robust, four-band SDWFS IRAC 
identifications have colors consistent with AGN, while the remainder
predominantly have colors consistent with galaxies at redshifts
above a few tenths.

Tasse \etal (2008a,b) have studied radio source properties on
the basis of {\it XMM-Newton}, CFHTLS, \SSS, and radio surveys in
the {\it XMM}-LSS field.  The radio surveys were carried out with
the Very Large Array at 74 and 325\,MHz, and with the Giant Meterwave
Radio Telescope (GMRT) at 230 and 610\,MHz.  Tasse \etal found
two classes of radio galaxies.  The first class
consists of massive galaxies that show no signs of infrared excess
due to a dusty torus.  These galaxies are preferentially found in cluster environments.
The second class consists of radio galaxies with less massive hosts.
These galaxies do show infrared torus emission and are located in large scale
underdensities.
These results are interpreted as being due to AGN that are fed via
two different types of accretion.  In this scheme, one mode (the
``quasar mode'') is radiatively efficient and arises from accretion
of cold gas onto a super-massive black hole.  The other mode (the
``radio mode'') is radiatively inefficient and results from accretion
of hot gas onto a super-massive black hole.  Quasar-mode objects
are predominantly located inside the Stern et al. (2005) wedge,
but radio-mode objects lie outside it.

Deeper (but lower spatial resolution) 1.4\,GHz observations of
approximately 7 deg$^2$ of the Bo\"otes field were presented by
de~Vries \etal (2002).  These data, obtained with the Westerbork
Synthesis Radio Telescope, reach a $5 \sigma$ sensitivity of 140
$\mu$Jy, approximately five times the depth of the FIRST survey,
and detect 3172 distinct sources.  At the 13\arcsec$\times$27\arcsec\,
resolution of the Westerbork beam, 316 sources are spatially resolved.
We consider only the 2856 unresolved radio sources.  The coarse
spatial resolution of this survey implies that a larger match radius
should be used than was used for FIRST.  However, the SDWFS data
are sufficiently sensitive that a larger match radius will yield a
prohibitively large number of spurious identifications.  We instead
opt for a 2\arcsec\ positional match radius where reliability should
be high but completeness will be compromised.  The 1392 (48.7\%)
IRAC counterparts thus conservatively identified are a poor probe
of the mid-IR identification rate for faint radio sources.
They probe are probe the mid-IR colors of radio
sources below the FIRST limit (Figure\,\ref{fig:xrayradio}).  As
expected (e.g., Seymour \etal 2008) at these fainter radio fluxes,
we see a large population of star-forming galaxies as shown by
their red [5.8]$-$[8.0] colors due to PAH emission in the reddest IRAC
bandpass.

\subsection{Mid-Infrared Detections of X-Ray Sources}

The XBo\"otes survey (Murray \etal 2005; Brand \etal 2006) 
imaged roughly 8.5 square degrees in Bo\"otes to a depth of 5\,ks
using the Advanced CCD Imaging Spectrometer (ACIS) on the {\it
Chandra X-ray Observatory}.  This survey, which consisted of 126
separate, contiguous ACIS observations, identified 3293 sources
detected with four or more counts in the full $0.5 - 7$\,keV band
(Kenter \etal 2005), corresponding to a flux of $7.8\times10^{-15}\,{\rm
erg}\, {\rm cm}^{-2}\, {\rm s}^{-1}$ for on-axis sources.  The
spurious detection rate was approximately 1\%.  The spatial resolution
of {\it Chandra} degrades quadratically with off-axis distance,
ranging from only 0\farcs6 (FWHM) on-axis to 6\farcs0 when 10\arcmin\
off-axis.  We considered the 2243 (68.1\%) of XBo\"otes sources
with positional uncertainties $\leq$ 2\arcsec\ and performed a
simple position match within a 3\arcsec\ match radius.  This match
radius will yield a 16.4\% spurious identification rate at the
$5\sigma$ limit of the 3.6\,$\mu$m data (4\arcsec\ diameter aperture), falling to 2.3\% at
8.0\,$\mu$m, assuming that IRAC sources are randomly distributed.
The spurious identification rate, of course, is lower for brighter
mid-IR sources.  After applying a small, empirically-derived shift
to the X-ray astrometric frame ($\Delta {\rm R.A.} = -0\farcs335,
\Delta {\rm Dec.} = +0\farcs467$), we find IRAC 3.6\,$\mu$m
identifications for 96.9\%\ of the subsample of unresolved XBo\"otes
sources with small positional accuracies; 80.0\% have robust,
four-band IRAC identifications.  Thus, while the generous match
radius would suggest a large number of spurious 3.6\,\micron\
identifications, the $>80\%$ match rate at 8.0\,$\mu$m suggests
that few of the shorter-wavelength identifications are, in fact,
incorrect.  Visual inspection shows that only $\sim$1\%\ of the
X-ray sources correspond to blank fields at 3.6\,$\mu$m, roughly
consistent with the expected spurious rate for the 4-count catalog.
Figure~\ref{fig:radioIDs} shows how the XBo\"otes identification
rate depends both on IRAC passband and mid-IR survey depth.
Relative to the original IRAC Shallow Survey, the deeper SDWFS data
nearly doubles the number of X-ray sources with robust, four-band
IRAC detections.

Figure~\ref{fig:xrayradio} shows the IRAC color-color properties
of XBo\"otes sources.  Similar to the results of Gorjian \etal
(2008), which matched XBo\"otes to the original IRAC Shallow Survey,
77.1\% of the X-ray sources satisfy the color-selection
criteria developed by Stern \etal (2005) to select active galaxies
in the mid-IR.  For comparison, Gorjian \etal (2008) found robust,
four-band IRAC identifications for only 42.9\% of the XBo\"otes
sources in the shallower IRAC Shallow Survey data, of which 65.2\%
have AGN-like IRAC colors.  If we restrict a similar analysis to
XBo\"otes sources with $\leq 2\arcsec$ positional accuracies at the
single-epoch $5 \sigma$ depths, 61.0\% have robust, four-band IRAC
identifications, of which 80.0\% satisfy the IRAC AGN selection
criteria of Stern \etal (2005).  In brief, then, the deeper SDWFS
data significantly increase the number of X-ray sources with robust,
four-band photometry.  However, the nature of the identified sources
does not appear to change significantly, with approximately 80\%
of the well-characterized X-ray sources still having mid-IR colors
indicative of AGN activity dominating at IRAC wavelengths.  As shown
by Hickox \etal (2007), stacking the X-ray data for IRAC-selected
AGN candidates lacking direct detections in the XBo\"otes data show
that such sources, on average, harbor high luminosity, obscured
AGN.

\subsection{Infrared-Variable Sources}
\label{ssec:irvar}

We have analyzed the four SDWFS epochs to perform a blind search for 
infrared-variable sources using difference imaging (e.g., Alard \& 
Lupton 1998; Wo{\'z}niak 2000).  Details and complete results will 
be reported elsewhere (Koz{\l}owski \etal 2009, in preparation) 
but we present initial findings here.  Each single-epoch mosaic
was prepared for difference-imaging by constructing a 
smoothed estimate of the {\it Spitzer} PSF.  We used Fourier methods to 
create new versions of the mosaics in which the native IRAC PSFs were
converted into Gaussians of similar width.  This removed the complex, 
rotated diffraction spikes for which difference imaging algorithms are not 
generally designed.   We then used the Difference Image Analysis 
package (DIA; Wo{\'z}niak 2000) to subtract successive SDWFS epochs.

The variability (dispersion) was estimated for each object detected in
the four difference images, where the average image of the four epochs 
was subtracted from each epoch.  The difference imaging techniques were 
very useful in the data processing phase discussed in \S2.2 because 
they allowed rapid identification of artifacts in the SDWFS mosaics.  
With almost all the sources vanishing (because they are not variable), 
even low level artifacts stood out.  More importantly, this technique
directly addresses the greatest limitation of mid-IR color selection of AGN:
that it begins to fail
as the AGN luminosities diminish to levels comparable to those of the host 
galaxies and the combined mid-IR colors become bluer (Fig.~\ref{fig:colors}).  
This makes it difficult to identify lower-redshift, lower-luminosity AGN.   
Time variability and difference imaging eliminate the contaminating hosts 
and extend the detectability of AGN in the mid-IR into the host-dominated 
regime where many X-ray and radio sources are also observed.

The overall dependence of variability on magnitude is shown in 
Figure~\ref{fig:variables}.  Candidate variable sources were selected as 
significant outliers from the general trend, 
and, unsurprisingly, the resulting sample 
is dominated by AGN with a distribution in mid-IR color-color space similar to 
the X-ray sources of Figure~\ref{fig:xrayradio}.
We calculated the dispersions separately at 3.6 and 4.5\,\micron, as
well as the correlations in the changes between the two bands.  
We designated an object as IR-variable if it changed by 
$\ge1.5 \sigma$ in both bands, if those changes were in synchrony 
(correlation $>0.5$), and if the uncertainties in the 3.6 and 4.5\,\micron\
magnitudes were below 0.05\,mag.   We identified 379 variable objects 
on the basis of these criteria, or about 45\,deg$^{-2}$ to the depth of a
single SDWFS epoch.   Of these, 272 objects were found to reside in the 
AGN wedge and 107 objects fell outside it (Figure~\ref{fig:variables}).

\subsection{High Proper Motion Sources}

Of the wide-area surveys done by {\it Spitzer} to date, SDWFS is
unique in its ability to identify high proper-motion mid-IR sources.
This is particularly promising for identifying the coolest brown
dwarfs, since such sources are very faint optically but bright in
the mid-IR.  For example, mid-T brown dwarfs have
$r-{\rm [3.6]} \approx 10$ (Strauss \etal 1999;
Tsvetanov \etal 2000; Patten \etal 2006),
implying that the single-epoch SDWFS mosaics reach the equivalent
of $r \approx 29$ for the coolest substellar sources.  
Thus, although SDWFS covers only a small fraction
of the sky, it samples $\approx 250$ times the volume of the
full-sky proper motion catalog of Luyten (1979) for mid-T brown dwarfs.  
Any cool brown dwarfs identified in a shallow, wide-area
survey such as SDWFS would be relatively nearby and thus might be
identifiable from their high proper motions alone.  In particular,
the abundances and mid-IR properties of the coldest brown dwarfs
(the so-called ``{\it Y dwarfs}'') are uncertain (or, at least,
untested) currently, providing additional strong motivation for
high proper motion mid-IR searches within the SDWFS survey.  Since
cool brown dwarfs have red [3.6]$-$[4.5] colors, we searched for
high proper motion objects independently in the two bluest IRAC
bands.

We began with the SExtractor catalogs for the four single-epoch IRAC
images at both 3.6 and 4.5~$\mu$m.  Each epoch contains approximately
500,000 identified sources, and we cross-matched the catalogs using
{\tt mergecats}, a routine from the {\tt imcat} toolkit written by
N.  Kaiser for rapid positional cross-correlations of large data
sets.  We matched the first-epoch SDWFS catalog (e.g., our reprocessed
original IRAC Shallow Survey) to both the third and fourth epochs
of SDWFS using a 10\arcsec\, search radius.  These last two SDWFS 
epochs were obtained approximately one month apart in Spring 2008,
a little more than four years after the first epoch (UT 2004 January
10$-$14).  
We then selected objects which moved by at least 1\farcs25 ($\approx
3.5\sigma$) between both epochs one and three, {\em and} between epochs
one and four.  Because epochs three and four were only a month apart
(Table~\ref{tbl.epochs}), this removed many spurious sources with
negligible impact on the highest proper motion sources.  Epoch two
was ignored for this preliminary investigation of high proper motion
SDWFS sources.  To remove additional spurious sources (e.g., cosmic
rays), we also required that the candidates were not saturated and
that they varied by less than 0.2\,mag between 2004 and 2008.  For
the 3.6\,\micron\ selection, we required that sources were brighter
than 18.5 mag in epoch~one (aperture-corrected 3\arcsec\ diameter
magnitude).  The corresponding requirement for the 4.5\,micron\
selection was [4.5]$\leq 18.0$.  Although these requirements are
somewhat conservative, they should identify the most reliable,
highest proper motion sources across the SDWFS data set.  Other
techniques, such as image subtraction and creating local reference
frames, should identify objects with substantially smaller proper
motions.

The criteria above led to samples of 27 3.6\,\micron-selected candidates
and 60 4.5\,\micron-selected candidates.  We visually inspected all
candidates.  Most proved spurious, with the primary contaminant
being merged, confused sources.  A few others were artifacts from
cosmic rays, the stellar bleed trails of bright stars, and sources
near the edge of the IRAC mosaics that survived the automated
culling.  After inspection, four sources survived, with the two
brightest identified independently at both 3.6 and 4.5~$\mu$m
and two fainter sources only identified at 4.5\,$\mu$m.  
Of the two single-band candidates, one omission
(SDWFS~J142534.1+325204) is attributable to a stellar bleed
trail affecting the source in the third-epoch 3.6\,$\mu$m data.  
The other source, SDWFS~J142723.4+330403, had 3.6~$\mu$m motions just
below our selection thresholds.  Because this source has very red
[3.6]$-$[4.5] colors, the 4.5\,$\mu$m data actually provide higher
signal-to-noise, and we consider this source robustly identified.
Table~6 presents the basic information for all four candidates, and
Figure~16 presents their $I$-band and multi-epoch IRAC images. 

The two brightest proper motion sources are also very bright at 
visible wavelengths and have [3.6]$-$[4.5] $\approx 0$, consistent 
with the Rayleigh-Jeans fall-off of main-sequence stars.  This 
inference appears to be borne out in the literature.
One of these bright objects, 
J143315.5+330838, is listed in version 2.3.2 of
the Guide Star Catalog as having $\mu$(RA) 
$=-70 \pm 3$\,mas\,yr$^{-1}$ and $\mu$(Dec) $=-324 \pm1$\,mas\,yr$^{-1}$
(Lasker \etal 2008), 
reassuringly consistent with our proper motion measurements (Table~6).
On the basis of its SDSS spectrum, West \etal (2008) assign this object
a spectral type of M5 -- again consistent with the type inferred
on the basis of the IRAC color.
The other bright IRAC high proper motion source, J142714.2+335959, 
has a SDWFS proper motion of
$\mu$(RA) $=-310$\,mas\,yr$^{-1}$ and $\mu$(Dec) $=-210$\,mas\,yr$^{-1}$.
This is quite close to the measured values of
$\mu$(RA) $=-288\pm3$\,mas\,yr$^{-1}$ and $\mu$(Dec) $=-210\pm2$\,mas\,yr$^{-1}$
given by Lepine \& Shara (2005; the LPSM-NORTH catalog).
The SDSS colors of this object appear consistent with it being of
spectral type M0 or perhaps M1 (Bochanski \etal 2007).

By contrast,
the two fainter SDWFS sources are very faint optically and have
red colors across the first two IRAC bands.  Their colors 
([3.6]$-$[4.5] $> 0.5$) imply mid-T brown dwarf spectral types (Patten
\etal 2006).  In particular, SDWFS~J142723.4+330403 has [3.6]$-$[4.5]=1.33,
implying a spectral type of approximately T7.  Eisenhardt et al.
(2009) reported on color-selected brown dwarf candidates with
[3.6]$-$[4.5]$>$1.5, implying spectral types later than T7.  Based
on estimated spectral types, the high proper motion sources are at
$\approx 100$\,pc, and have space velocities of order 150\,km\,s$^{-1}$ 
relative to the Sun.

\subsection{SDWFS at the Depth of WISE}

The {\sl Wide-field Infrared Survey Explorer (WISE)} is scheduled to launch
in 2009 November.  {\sl WISE} will survey the full sky with a 
cryogenically-cooled 40\,cm telescope in four passbands at 3.3, 4.7, 12, and 23\,$\mu$m.
The two shortest {\sl WISE} passbands are 
similar to the two shortest IRAC passbands.  {\sl WISE} will reach $5 \sigma$
point source depths of 120 and 160~$\mu$Jy at 3.3 and 4.7\,$\mu$m, respectively. 

Figure~\ref{fig:WISEdepth} shows the colors of SDWFS sources that would be
detected by {\sl WISE} above ${\rm [3.6]} = 15.91$ and ${\rm [4.5]} = 15.12$\,mag.  

At these depths, SDWFS identifies 11,910 non-saturated sources with robust, 
two-band detections.  Because {\sl WISE} does not have equivalents of 
the IRAC 5.8 and 8.0\,$\mu$m detectors, one might expect that 
{\sl WISE} will be challenged to separate AGN from galaxies using mid-IR color criteria
alone.  However, as is clear from Figure~\ref{fig:WISEdepth}, at the shallow depth
of the planned {\sl WISE} survey, the ${\rm [5.8]} - {\rm [8.0]}$ 
color is unnecessary: nearly every source redder than 
${\rm [3.6]} - {\rm [4.5]} = 0.5$ is an AGN.  The handful of sources to 
the right of the AGN wedge that lie above this simple color cut 
are presumably star-forming galaxies at redshifts of a few tenths with 
dust emission producing their red IRAC colors.  Such sources are easily 
discriminated based on their optical properties.

Based on the 4\arcsec\ diameter aperture photometry from the 
v3.4 public-release SDWFS catalogs, {\sl WISE}
will detect $\sim90$ sources deg$^{-2}$ with 
${\rm [3.6]} - {\rm [4.5]} \geq 0.5$, of which 90\% will reside within 
the Stern et al. (2005) AGN wedge.  Similarly, 90\% of sources residing within
the Stern et al. (2005) wedge will have 
${\rm [3.6]} - {\rm [4.5]} \geq 0.5$.  
For comparison, SDSS finds 15 quasars deg$^{-2}$ to $i^* = 19.1$ 
(Richards \etal 2002).  It is thus expected that 
{\sl WISE} will be a very powerful instrument for studying the 
cosmic evolution of AGN with minimal bias due to obscuration.

\section{Summary}

\SSS/IRAC imaging of 10 deg$^2$ in Bo\"otes has detected over half of a 
million infrared sources as part of the \SSS\  Deep, Wide-Field Survey (SDWFS).
Images and catalogs containing the four-band IRAC photometry for these sources
are now available online for the community to use.  It is anticipated
that SDWFS will facilitate a wide variety of extragalactic science;
some of the avenues already being explored have been described above.  
A forthcoming paper (Eisenhardt et al. 2009, in preparation) details the 
discovery of ultracool brown dwarf candidates from the SDWFS data
set, including the likely identification of T7 and T8 field brown dwarfs.
Other projects that are also being actively developed have been described, 
including a comprehensive study of infrared-variable AGN 
(Koz{\l}owski et al. 2009, in preparation), and an investigation of infrared background fluctuations
(Cooray et al. 2009, in preparation).  These data will soon become
publicly available in a format that combines the SDWFS photometry with 
optical photometry from the NDWFS (Jannuzi \& Dey 1999).  The combined
catalogs will facilitate many entirely new applications of the dataset 
from photometric redshift studies to large-scale structure.

\acknowledgments

This work is based on observations made with the {\it Spitzer Space
Telescope}, which is operated by the Jet Propulsion Laboratory,
California Institute of Technology under contract with the National 
Aeronautics and Space Administration (NASA).  Support for this 
work was provided by NASA through award number 1314516 
issued by JPL/Caltech.  This research made use of Montage, funded by the
NASA's's Earth Science Technology Office, Computation Technologies Project, under Cooperative
Agreement Number NCC5-626 between NASA and the California Institute
of Technology.  Montage is maintained by the NASA/IPAC Infrared
Science Archive.  IRAF is distributed by the National Optical Astronomy 
Observatory, which is operated by the Association of Universities for 
Research in Astronomy (AURA) under cooperative agreement with the 
National Science Foundation.  Support for MB was provided by the W. M. 
Keck Foundation.  The authors wish to thank Andy Gould, whose suggestions
improved the manuscript.

Facilities:  \facility{{\it Spitzer Space Telescope} (IRAC)}

{}

\begin{figure*}
\epsscale{0.65}
\plotone{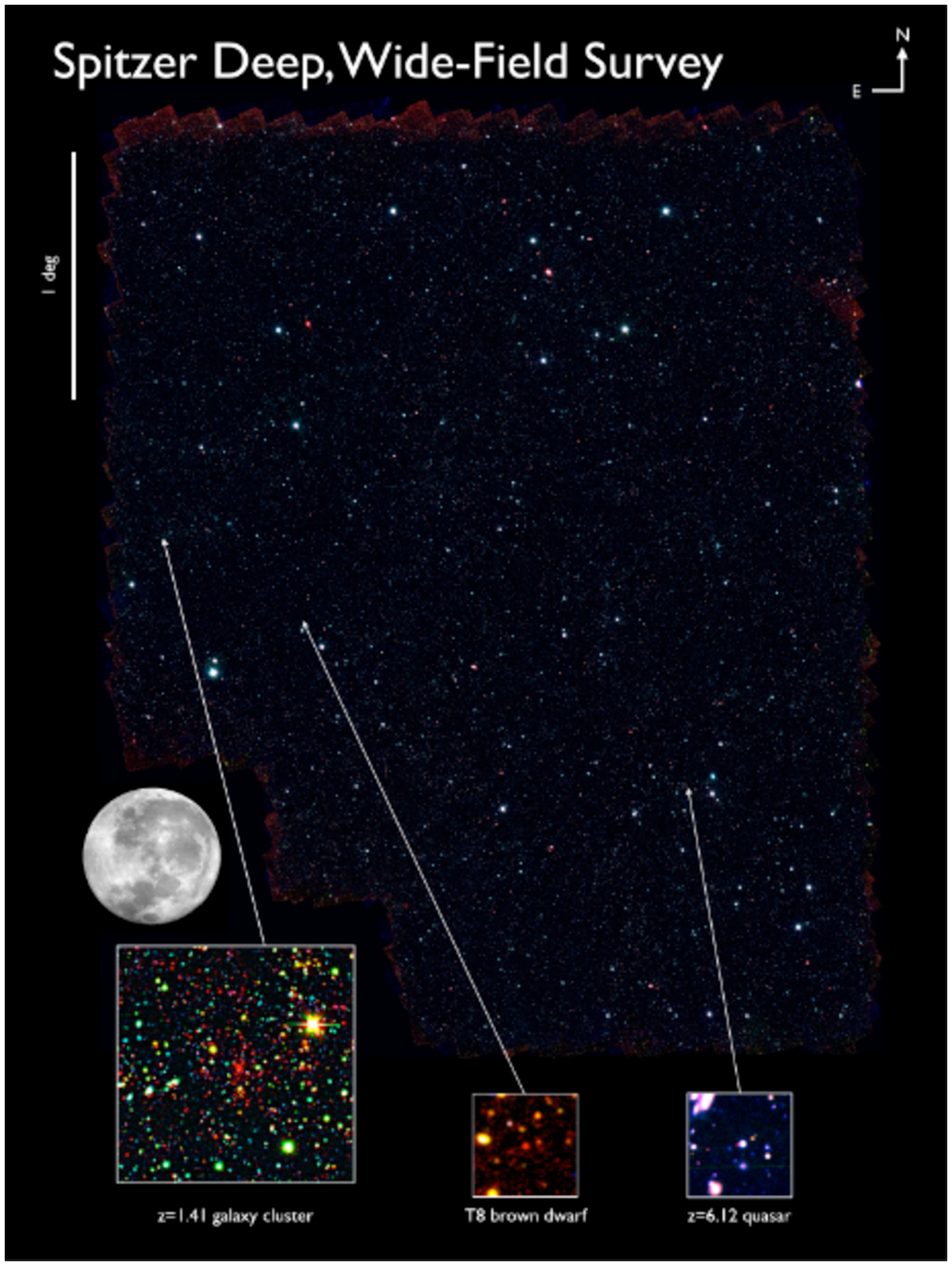}
\caption{
False-color IRAC image of the full SDWFS field using the full, 
four-epoch dataset (RGB = 8.0, 4.5, 3.6\,\micron).  The 
full Moon is shown to indicate the angular scale.  Three interesting targets 
identified from their IRAC 
properties are indicated:  a spectroscopically confirmed galaxy cluster 
at $z = 1.41$ (Stanford et al. 2005; $7^\prime\times7^\prime$; 
RGB = 4.5\,\micron, $I, B_{\rm W}$), a very late-type, field brown dwarf 
(Eisenhardt et al. 2009, in preparation; $1^\prime\times1^\prime$; 
RGB = 4.5, 3.6\,\micron, $J$), and a radio-loud quasar at $z = 6.12$ 
(Stern \etal 2007; McGreer \etal 2006; $1^\prime\times1^\prime$; RGB = $I, R, B_{\rm W}$).  
The galaxy cluster was the most distant known when first identified 
from the first epoch of SDWFS data.  The brown dwarf is amongst the ten 
coldest known.  The quasar was the first non-SDSS quasar identified 
at $z > 6$ and is the most distant radio-loud source currently known.
\label{fig.field}}
\end{figure*}

\begin{figure}
\plotone{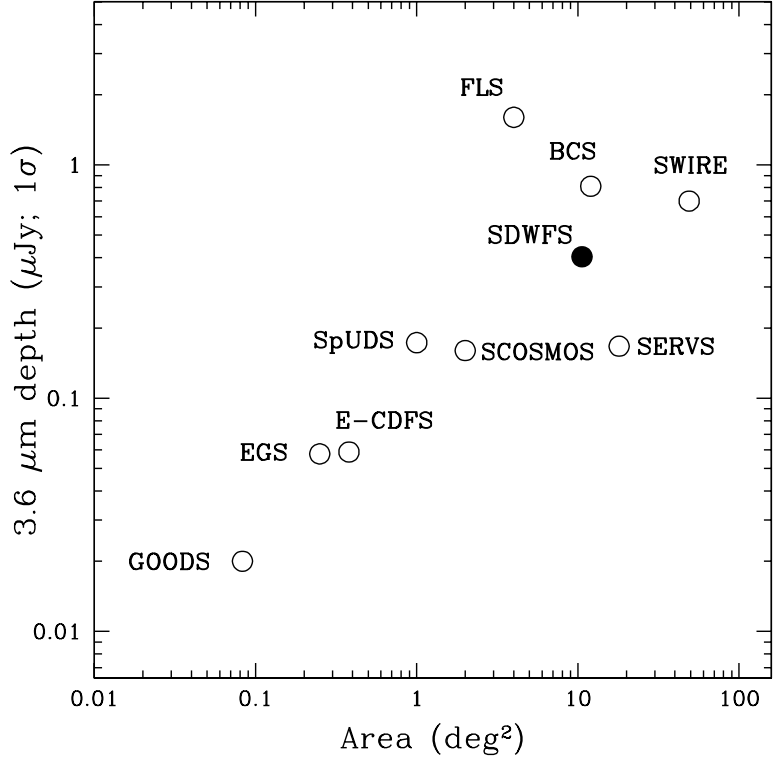}
\caption{Comparison of SDWFS 3.6\,\micron\ depth and total area to
the major \SSS/IRAC extragalactic surveys, including GOODS (Great Observatories
Origins Deep Survey), EGS (Extended Groth Strip), E-CDFS (Extended Chandra Deep
Field South), SpUDS (\SSS\ Public Legacy Survey of UKIDSS Ultra-Deep Survey),
SCOSMOS (\SSS\ Deep Survey of HST COSMOS 2-Degree ACS Field),
SERVS (\SSS\ Extragalactic Representative Volume Survey),
BCS (Blanco Cluster Survey), SWIRE (\SSS\ Wide-area Infrared
Extragalactic Survey), and FLS (\SSS\ First-Look Survey).
All depths are SENS-PET 1$\sigma$ point-source sensitivities calculated
under low background conditions, except for the FLS and SCOSMOS, which
assume medium backgrounds.  All surveys are contiguous except
SWIRE and SERVS, which are divided into six and five sub-fields,
respectively.
SDWFS is the only four-band, four-epoch
wide-area \SSS\ survey, thus enabling unique variability studies.
\label{fig.etendue}}
\end{figure}

\begin{figure*}
\epsscale{1.00}
\plotone{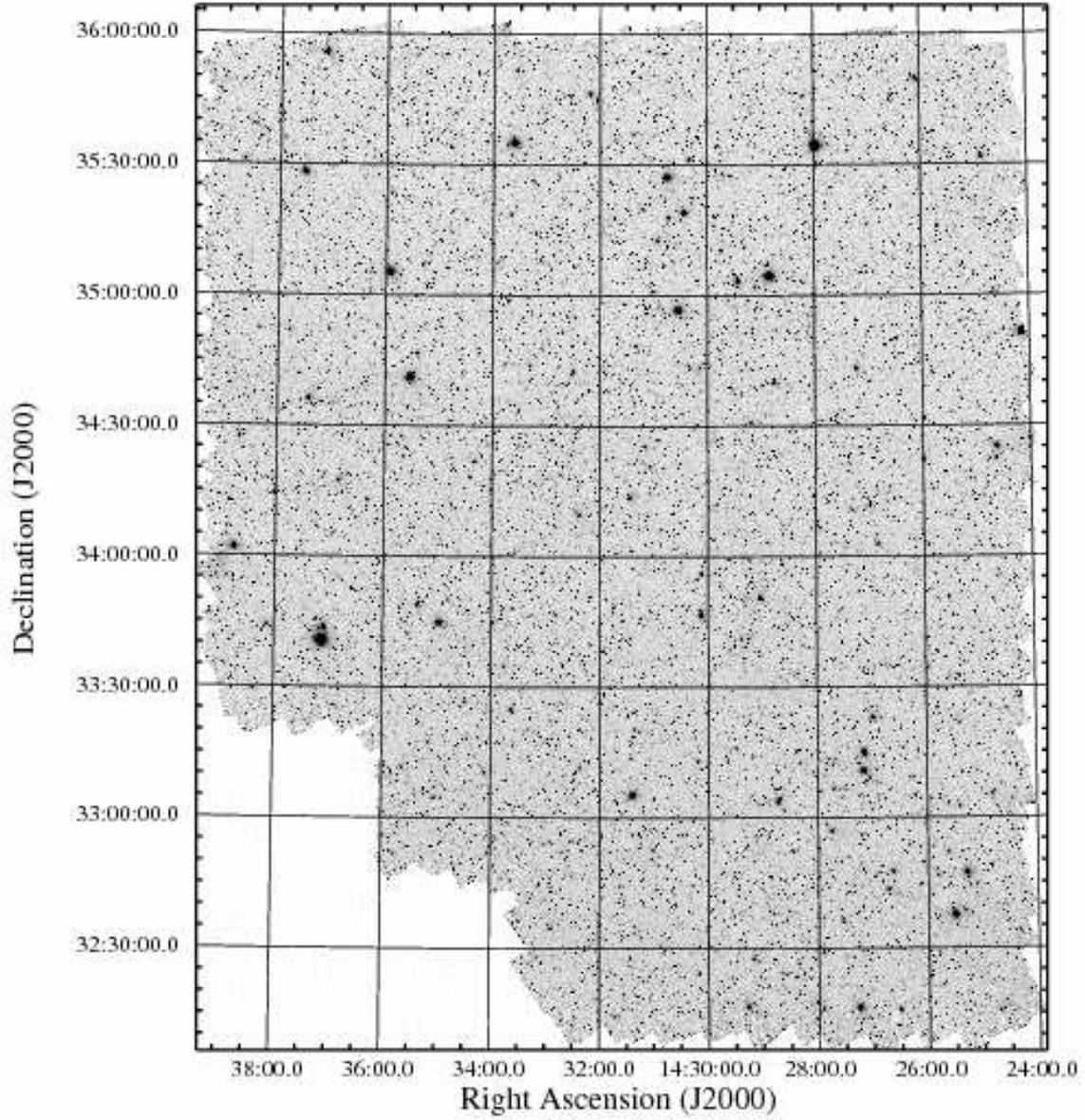}
\caption{Total SDWFS IRAC 3.6\,\micron\ coadd (four coadded epochs)
of the Bo\"otes field.
\label{ch1_image}}
\end{figure*}

\begin{figure*}
\plotone{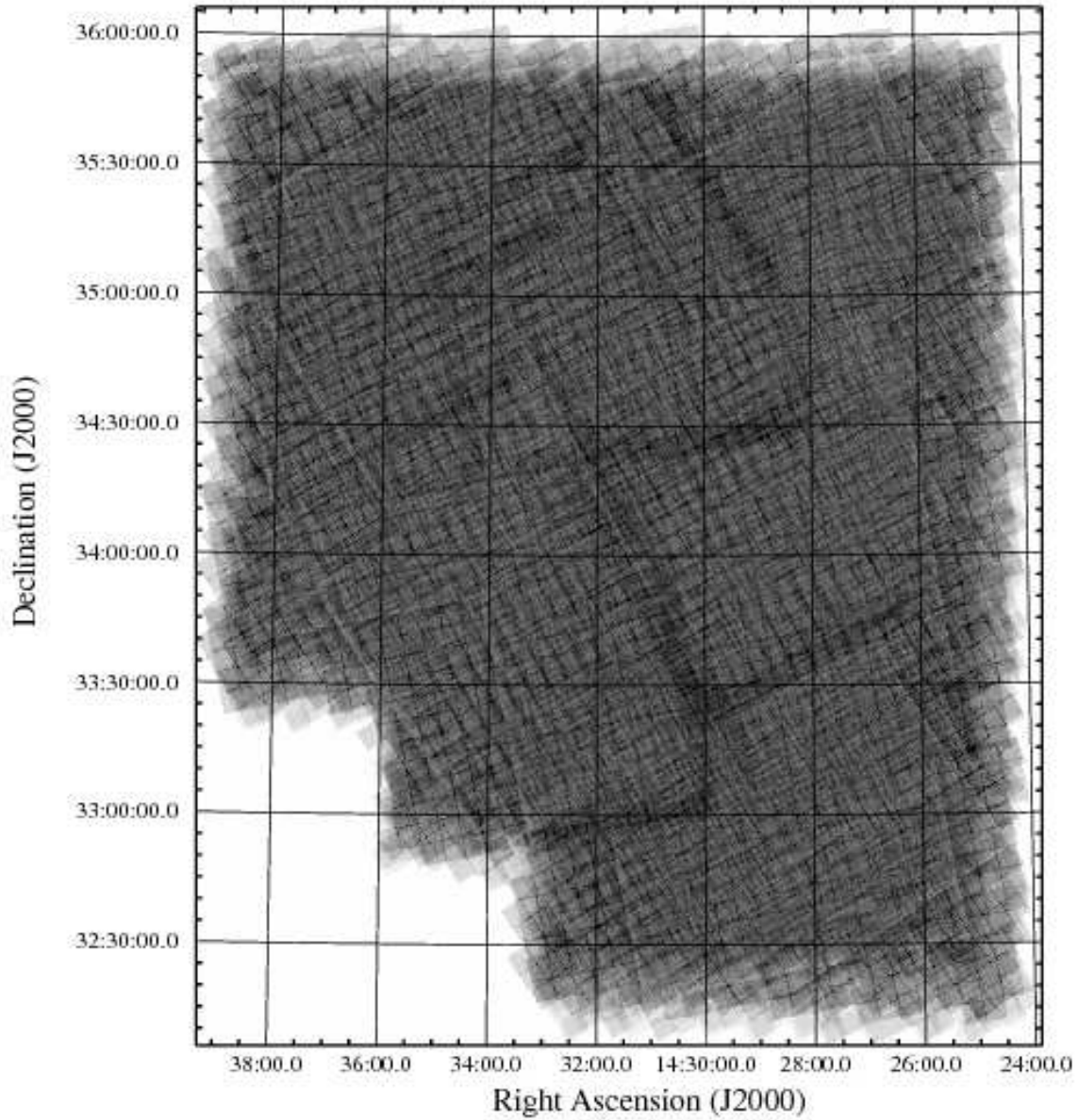}
\caption{Total SDWFS IRAC 3.6\,\micron\ coverage map (four coadded
epochs) illustrating the uniformity of depth over the Bo\"otes field.  
The coverage ranges from a maximum
of $19\times$30\,s (black) in the interior to no coverage (white), 
at the edge.  The total area
observed with at least one 30\,s exposure is 10.6 square degrees.
The total area covered by the other three IRAC bands is nearly identical,
but the coverage differs in detail at the edges of the field because 
the 3.6/5.8\,\micron\ and 4.5/8.0\,\micron\ fields of view are offset
from each other.  
\label{cvg_map}}
\end{figure*}

\begin{figure*}
\plotone{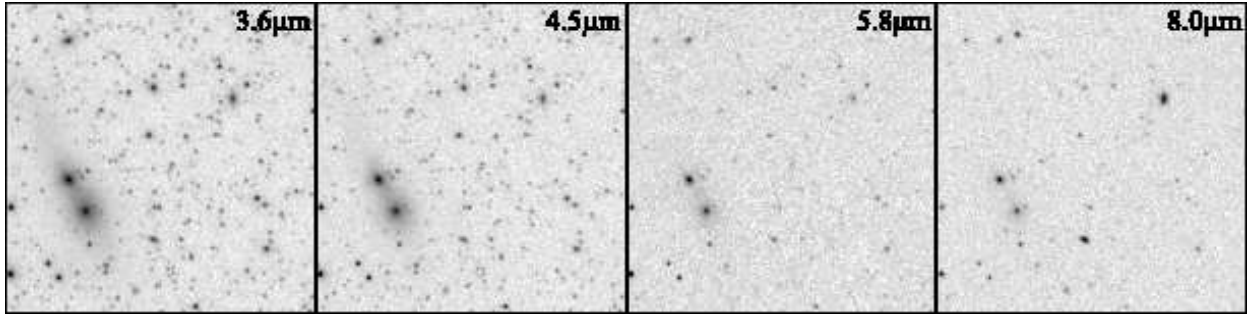}
\caption{Montage of cutouts from the four-epoch $12\times30$\,s
mosaics of the Bo\"otes survey field, illustrating the source
densities in the IRAC bands.  Coverage is roughly $5^\prime
\times5^\prime$, centered at $(\alpha,\delta)_{\rm J2000} =$
(14:31:48,+34:43:00).  North is up, and east is to the left.  The
two relatively bright galaxies in the east half of the frame are
an interacting pair of red elliptical galaxies.  They were presented
as evidence of ``dry merging'' in van Dokkum (2005; galaxies 17-596
and 17-681 in Figures~3a and ~4a).
\label{fig:cutouts}}
\end{figure*}

\begin{figure}
\plotone{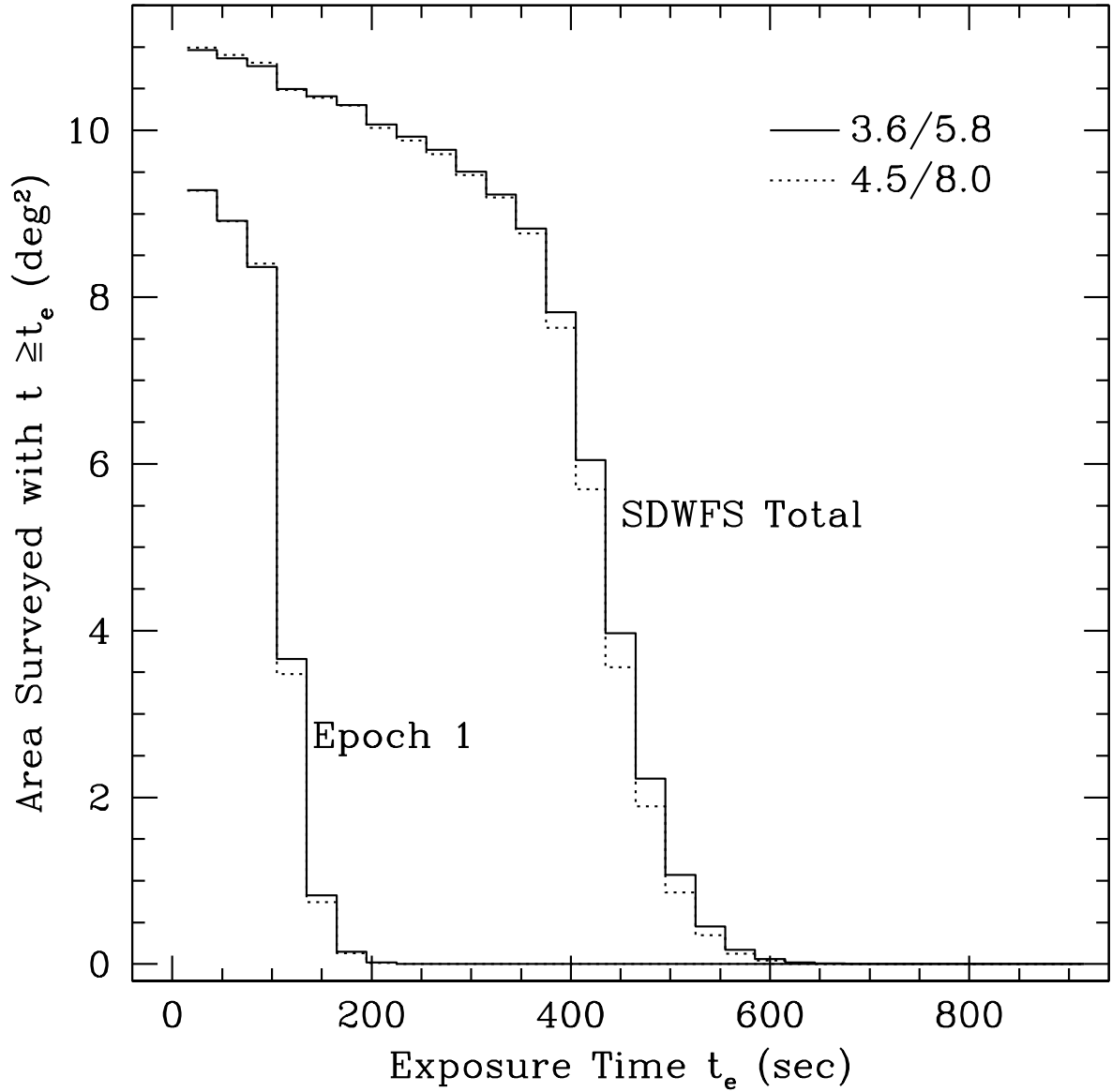}
\caption{Cumulative area coverage as a function of exposure time
for IRAC observations of the Bo\"otes survey field.   The median
coverage obtained in a single epoch and the full survey are about
90 and 420\,s, respectively, in all bands.  
\label{fig:coverage}}
\end{figure}

\begin{figure}
\plotone{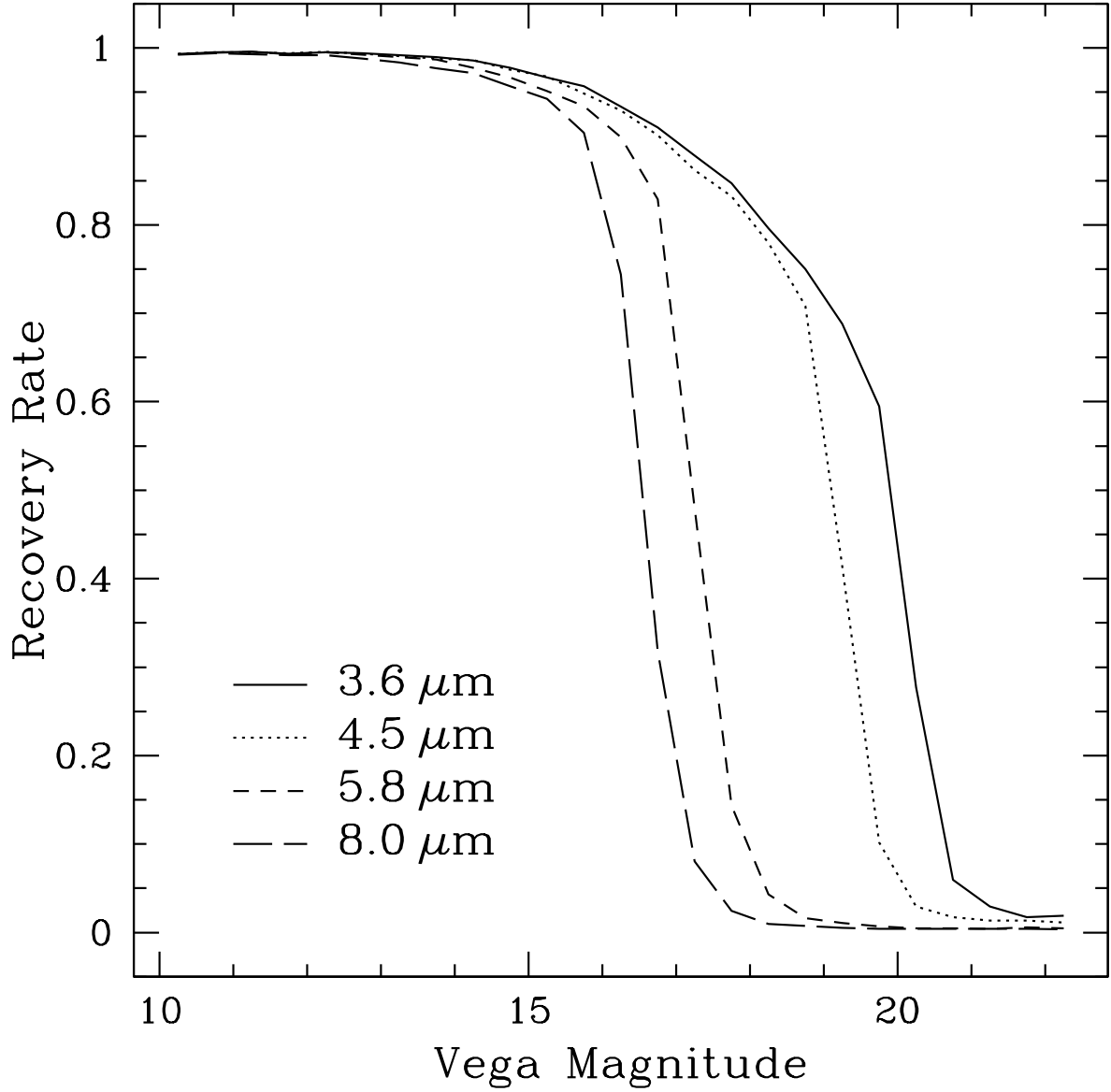}
\caption{Fraction of simulated point-sources recovered as a function
of input magnitude for the total coadd of all four IRAC observations
of the Bo\"otes survey field.
\label{brodwin_sims}}
\end{figure}

\begin{figure}
\plotone{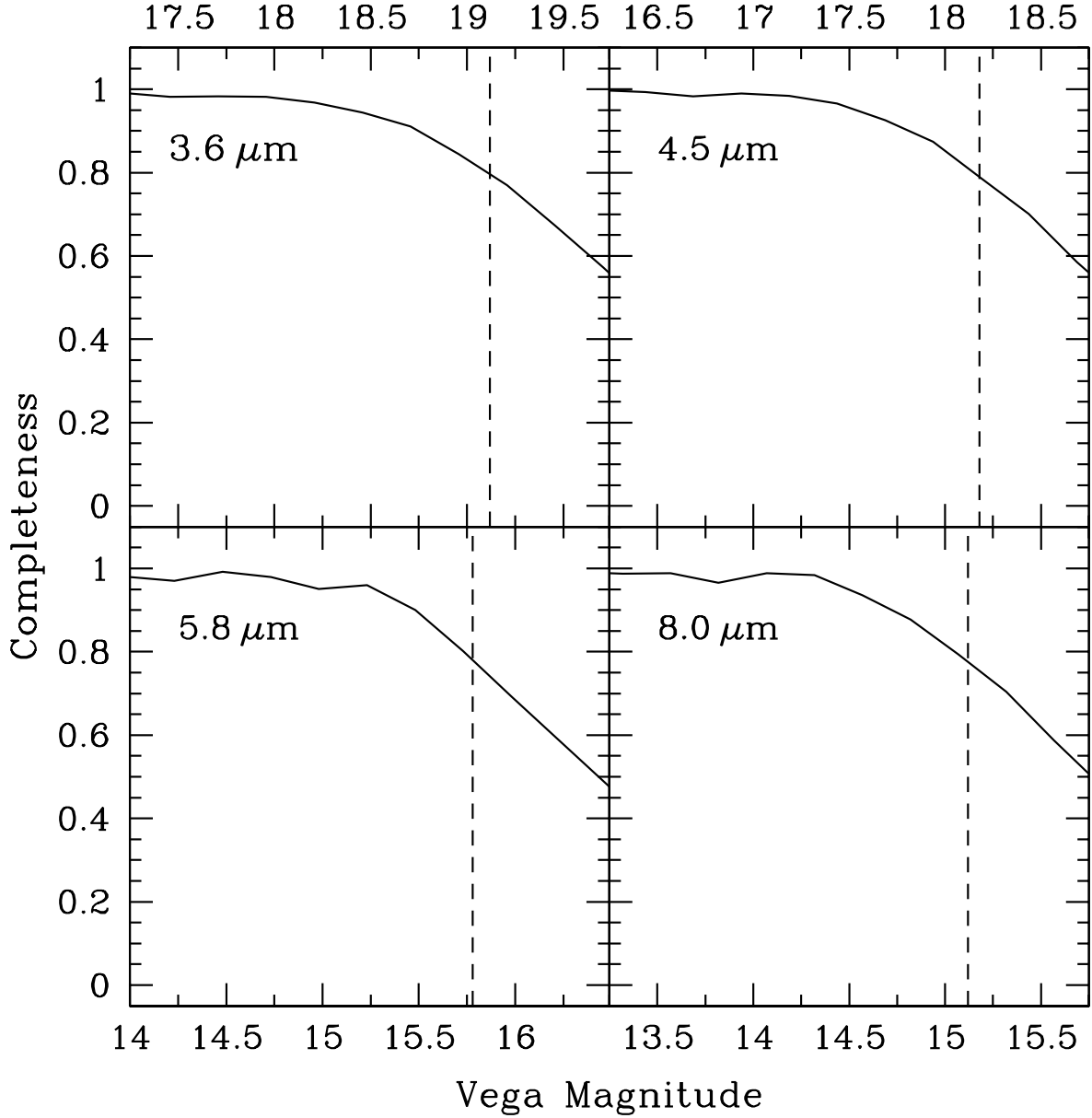}
\caption{Empirical single-epoch fractional completeness estimates for SDWFS.
Each panel shows the fraction of SDWFS sources identified in the
total $12\times30$\,s SDWFS coadds that are also found in the
epoch four $3\times30$\,s coadd.  The 5$\sigma$ sensitivity limits 
established with
the Monte Carlo approach are shown as vertical dashed lines.
\label{griffith1}}
\end{figure}
\subsection{Data Reduction}

\begin{figure}
\plotone{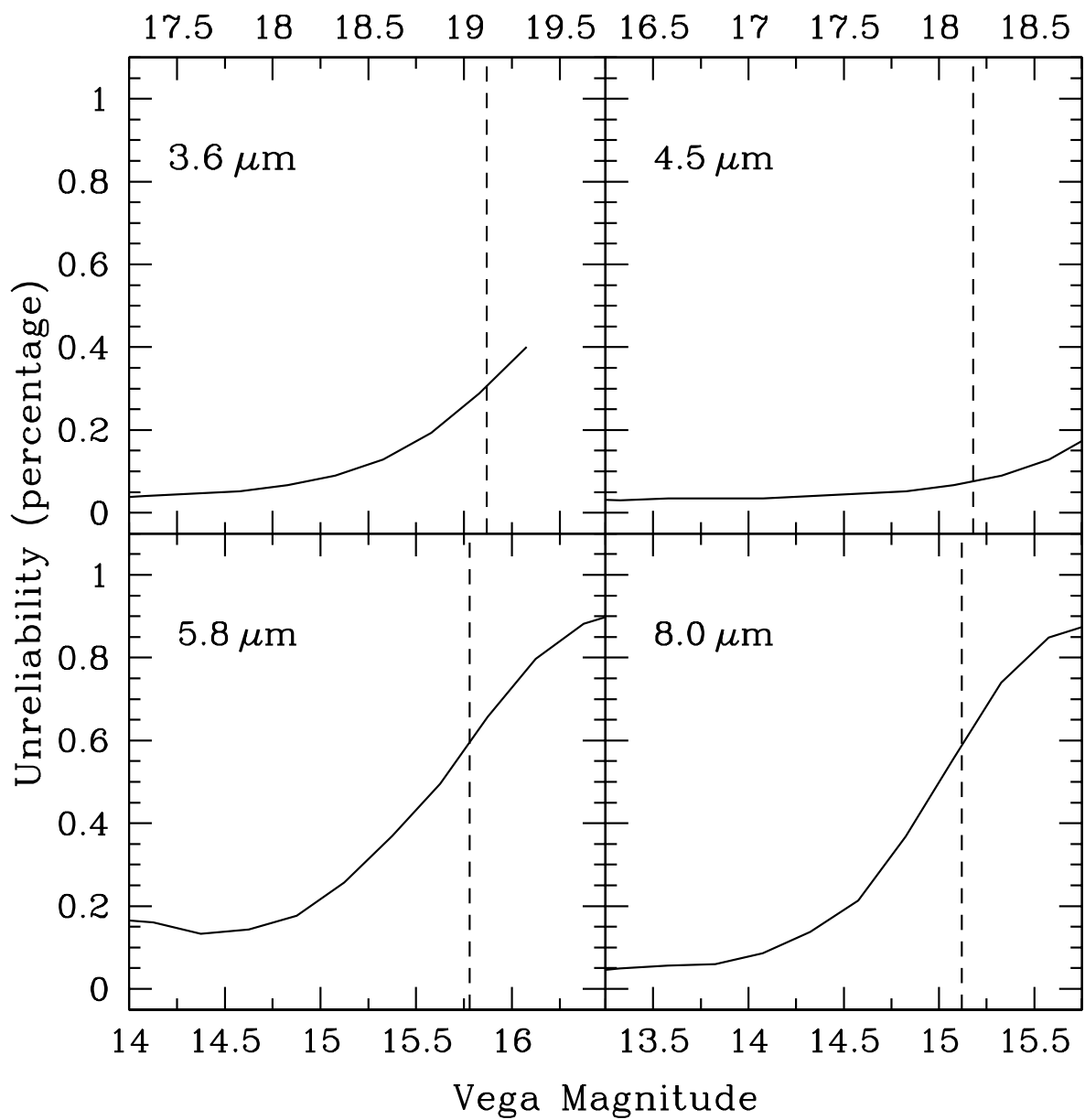}
\caption{Empirical single-epoch unreliability estimates for 
SDWFS.  Here, unreliability is defined as the percentage of sources detected at
$3\times30$\,s (in SDWFS epoch 4) that are {\sl not} found 
in the deeper $12\times30$\,s SDWFS coadd.
The 5$\sigma$ sensitivity limits established with
the Monte Carlo approach are shown as vertical dashed lines.
For sources brighter than the 5$\sigma$ single-epoch sensitivity limits, SDWFS
detections are at least 99.4\% reliable.
\label{griffith2}}
\end{figure}
\clearpage

\begin{figure}
\plotone{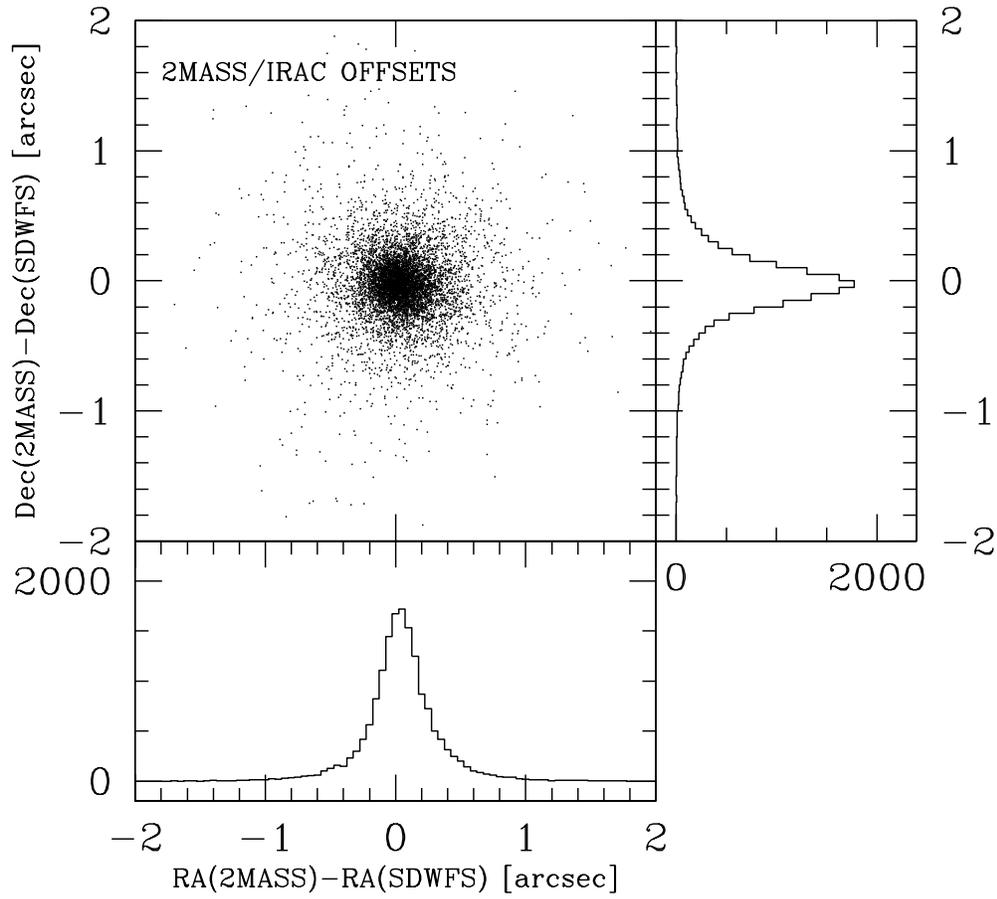}
\caption{Astrometric offsets (2MASS - SDWFS) between catalog positions 
drawn from the total IRAC 3.6\,\micron\ coadd and positions of all 
16,045 2MASS sources that lie in the 10.6 square degree region of 
Bo\"otes covered by SDWFS.  
\label{astrometry.2mass}}
\end{figure}

\begin{figure}
\plotone{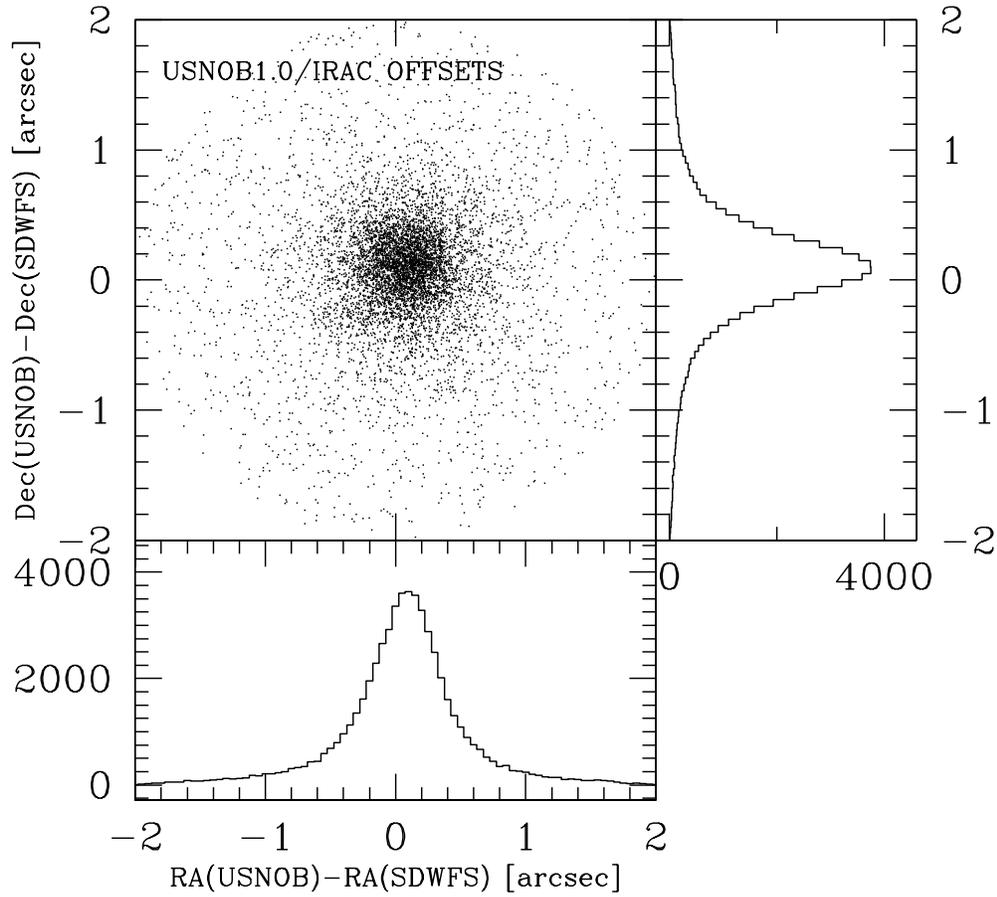}
\caption{Astrometric offsets (USNOB - SDWFS) between catalog positions drawn 
from the total SDWFS 3.6\,\micron\ coadd and positions of sources in the
USNOB1.0 catalog for the 10.6 square degree Bo\"otes field.  For
clarity, the central panel shows only 7500 of the $\sim56,700$ matches.
\label{astrometry.usnob}}
\end{figure}

\begin{figure*}
\plotone{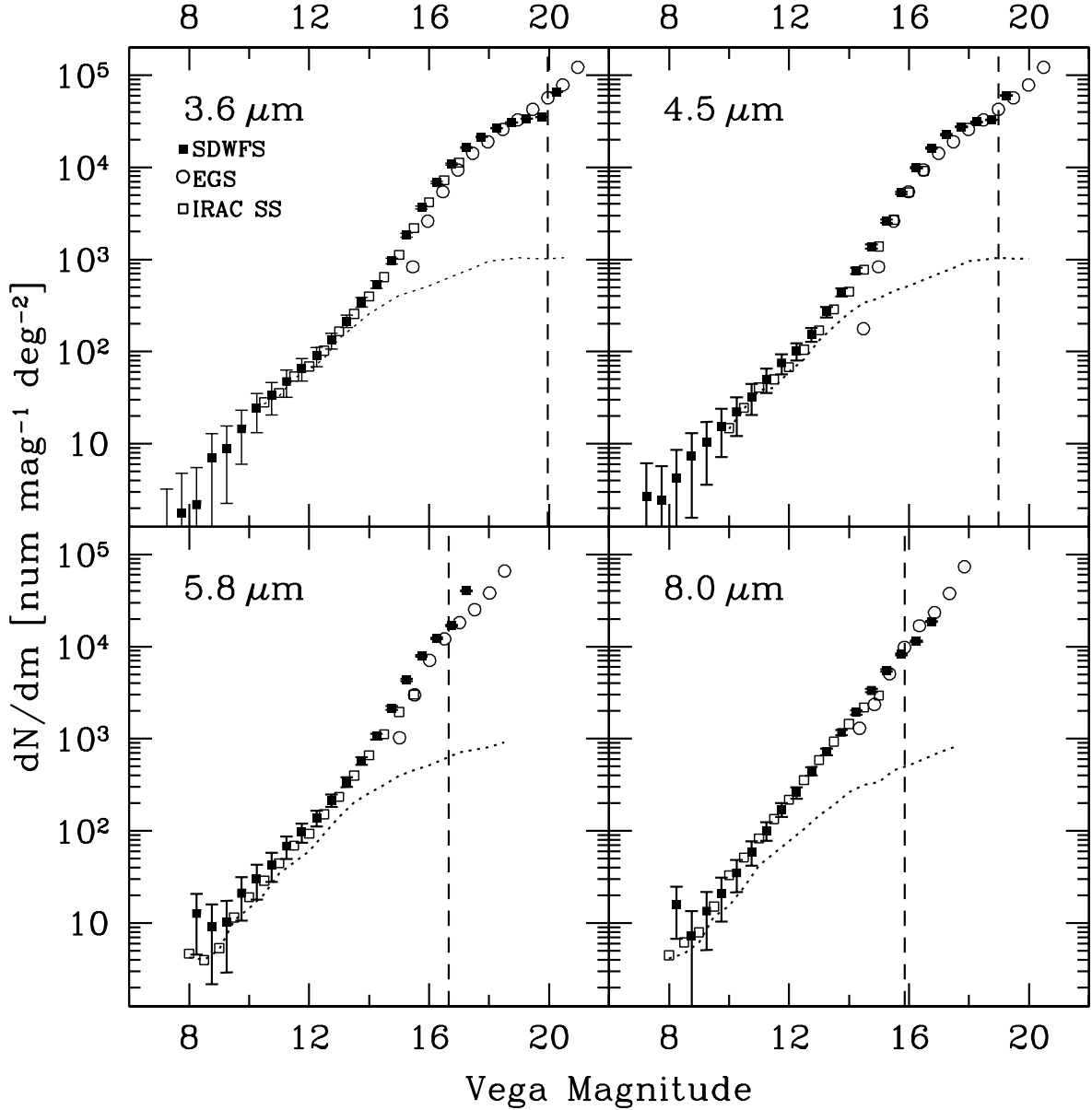}
\caption{
Differential source counts for SDWFS.  Solid squares show the total
counts (stars$+$galaxies) drawn from the full-depth SDWFS mosaic.
Open squares are the source counts originally measured for the IRAC
Shallow Survey (the first SDWFS epoch) by Fazio \etal (2004).  Open
circles indicate the counts tabulated in the deep IRAC survey of
the Extended Groth Strip by Barmby \etal (2008).  Dotted lines
indicate star counts corresponding to point-sources based on a 
Galactic model (tabulated by Fazio \etal 2004).  
All counts plotted have been corrected for
incompleteness using the results of the simulations 
described in \S~\ref{sec:completeness}.
The vertical dashed lines indicate the
5$\sigma$ sensitivity limits corresponding to 3\arcsec\ diameter
apertures (Table~\ref{tbl.depths}).
\label{fig:sdwfs_counts}}
\end{figure*}

\begin{figure}
\plotone{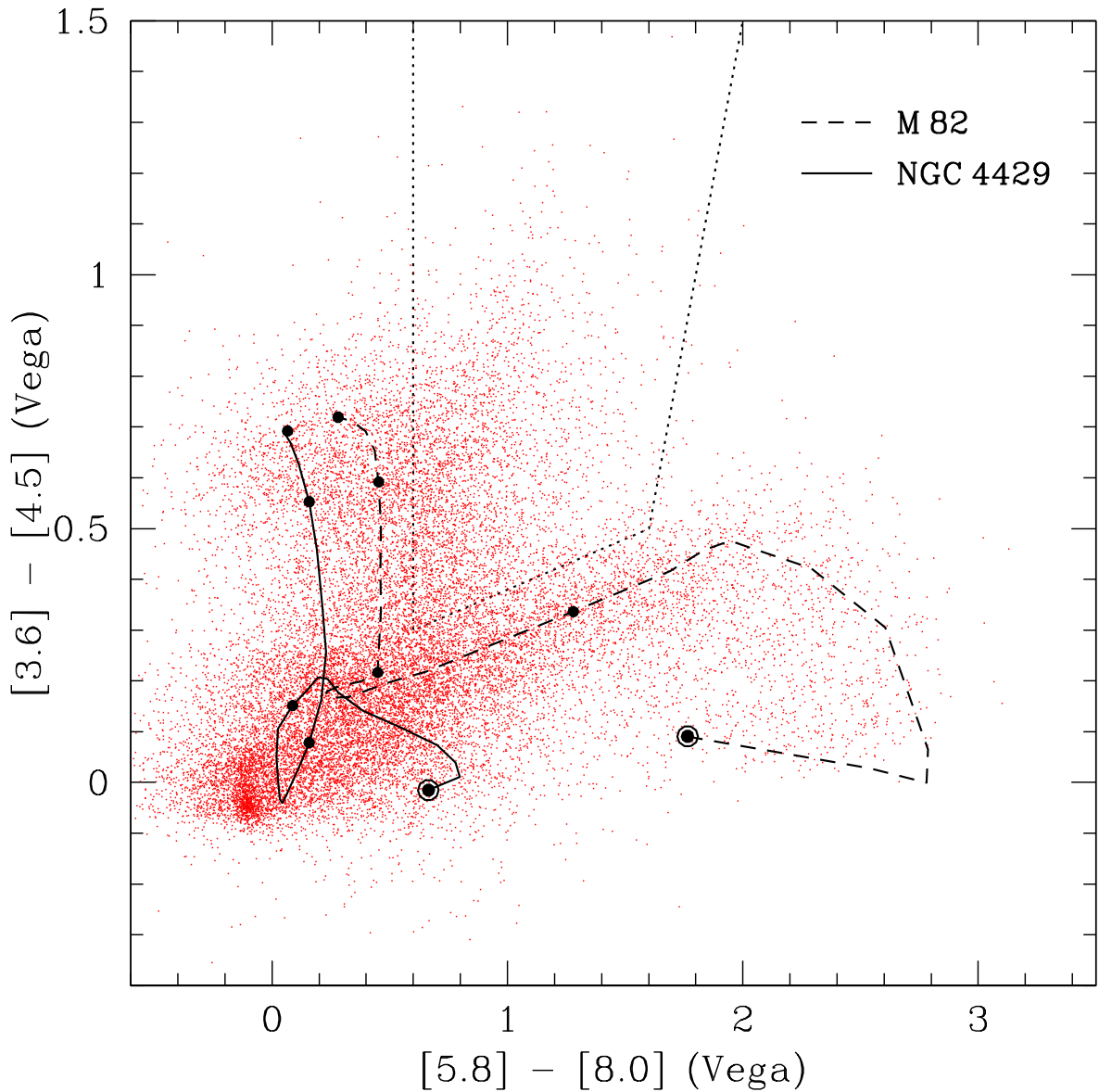}
\caption{IRAC color-color diagram for 20,000 randomly-selected SDWFS 
sources detected at $5\sigma$ significance or greater (red dots).
All photometry shown derives from 3\arcsec\ diameter
aperture magnitudes, corrected uniformly to 12 pixel radius magnitudes
(24\arcsec).  
The $z\le2$ color
tracks of two non-evolving galaxy templates (Devriendt \etal 1999)
are shown: the quiescent S0/Sa galaxy NGC\,4429 (solid line),
and the nearby starburst M82 (dashed line).  The solid dots 
correspond to $z=0.5$, 1.0, 1.5, and 2.0, and the dot-in-circles 
indicate $z=0$.  The dotted line defines a region of this color-color
space that is known to be occupied predominantly by active galaxies
(Stern \etal 2005).
\label{fig:colors}}
\end{figure}

\begin{figure}
\epsscale{1.00}
\plotone{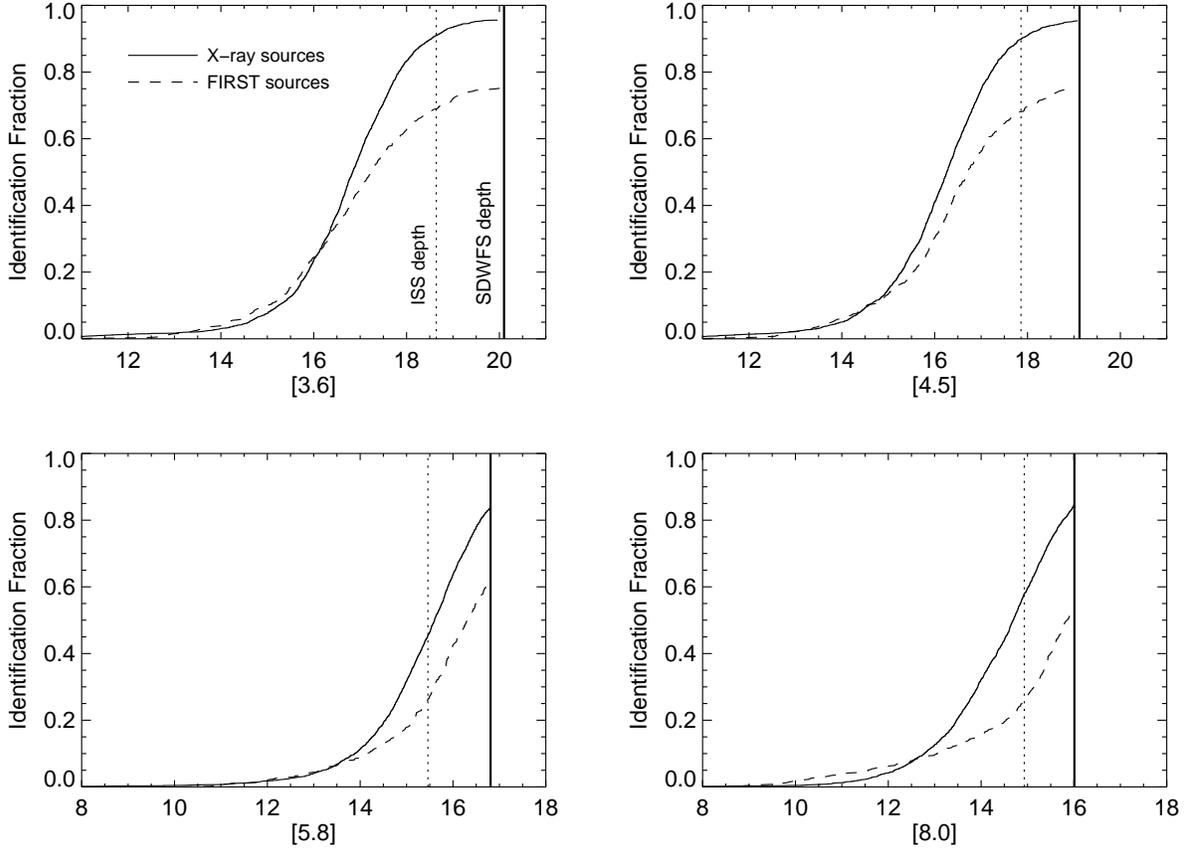}
\caption{Identification fraction of on-axis X-ray sources (solid
curve; $S_{\rm 0.5-7\, keV} \simgt 8 \times 10^{-15}\, {\rm erg}\,
{\rm cm}^{-2}\, {\rm s}^{-1}$) and isolated radio sources from the
FIRST survey (dashed curve; $S_{\rm 1.4\, GHz} \geq 0.75$~mJy) in
the Bo\"otes field for all four IRAC bands.  Vertical dotted lines
show the 5$\sigma$ depth of the original IRAC Shallow Survey
(Eisenhardt et al. 2004).  Vertical solid lines show the 5$\sigma$
depth of SDWFS.  Dotted curves show the spurious identification
rate as a function of magnitude for 2\arcsec\ and 3\arcsec\ match
radii.
\label{fig:radioIDs}}
\end{figure}

\begin{figure}
\epsscale{1.00}
\plotone{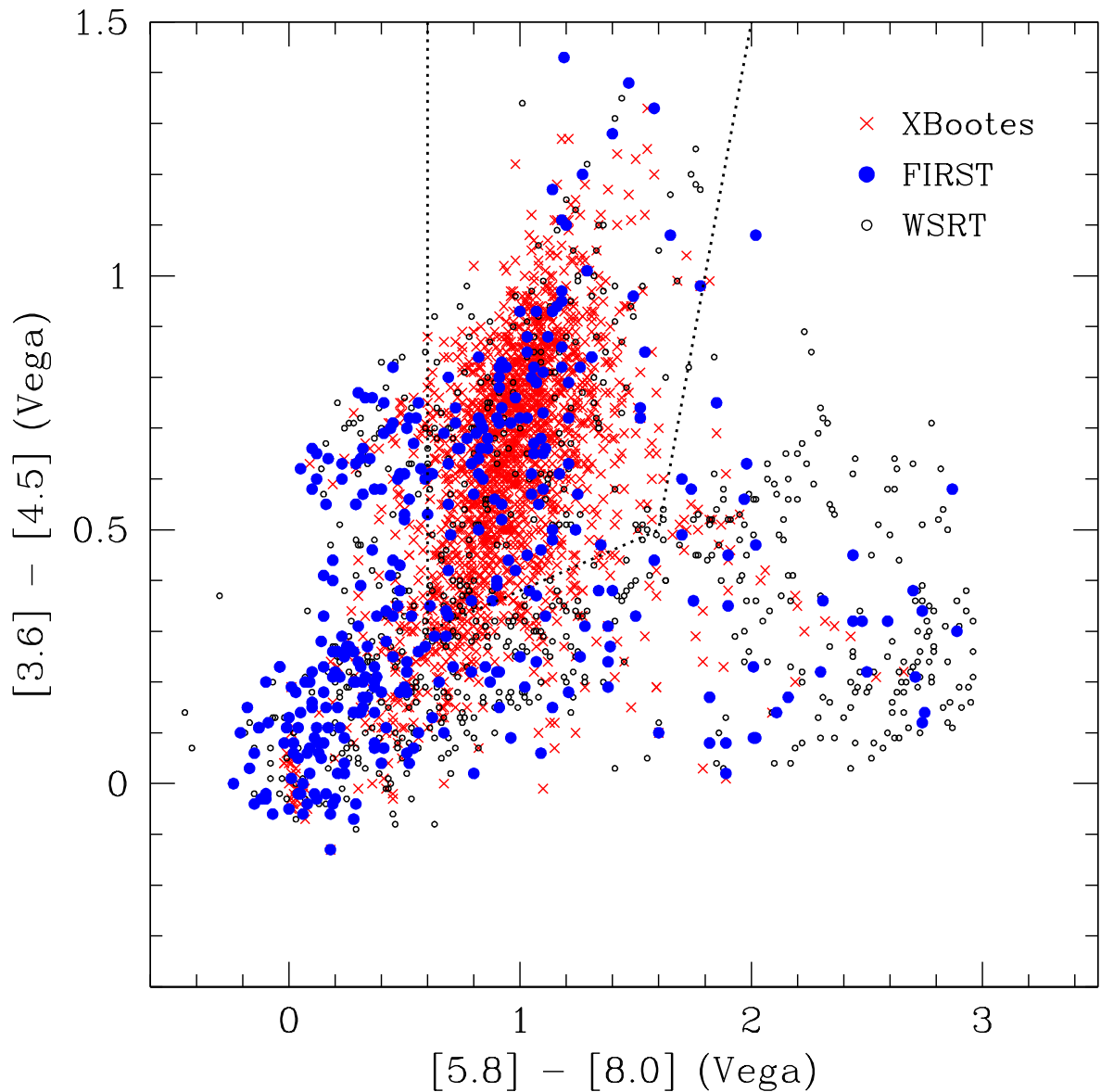}
\caption{IRAC color-color plot for X-ray sources (red crosses) and
radio sources (filled blue circles and open circles) detected
in all four IRAC bands.  Open circles
show 1.4~GHz sources fainter than the FIRST 0.75~mJy limit
identified from deeper Westerbork Synthesis Radio Telescope (WSRT)
observations (de Vries et al. 2002).  Dotted lines indicate the
(Stern et al. 2005) AGN wedge.
\label{fig:xrayradio}}
\end{figure}

\begin{figure*}
\epsscale{1.00}
\plotone{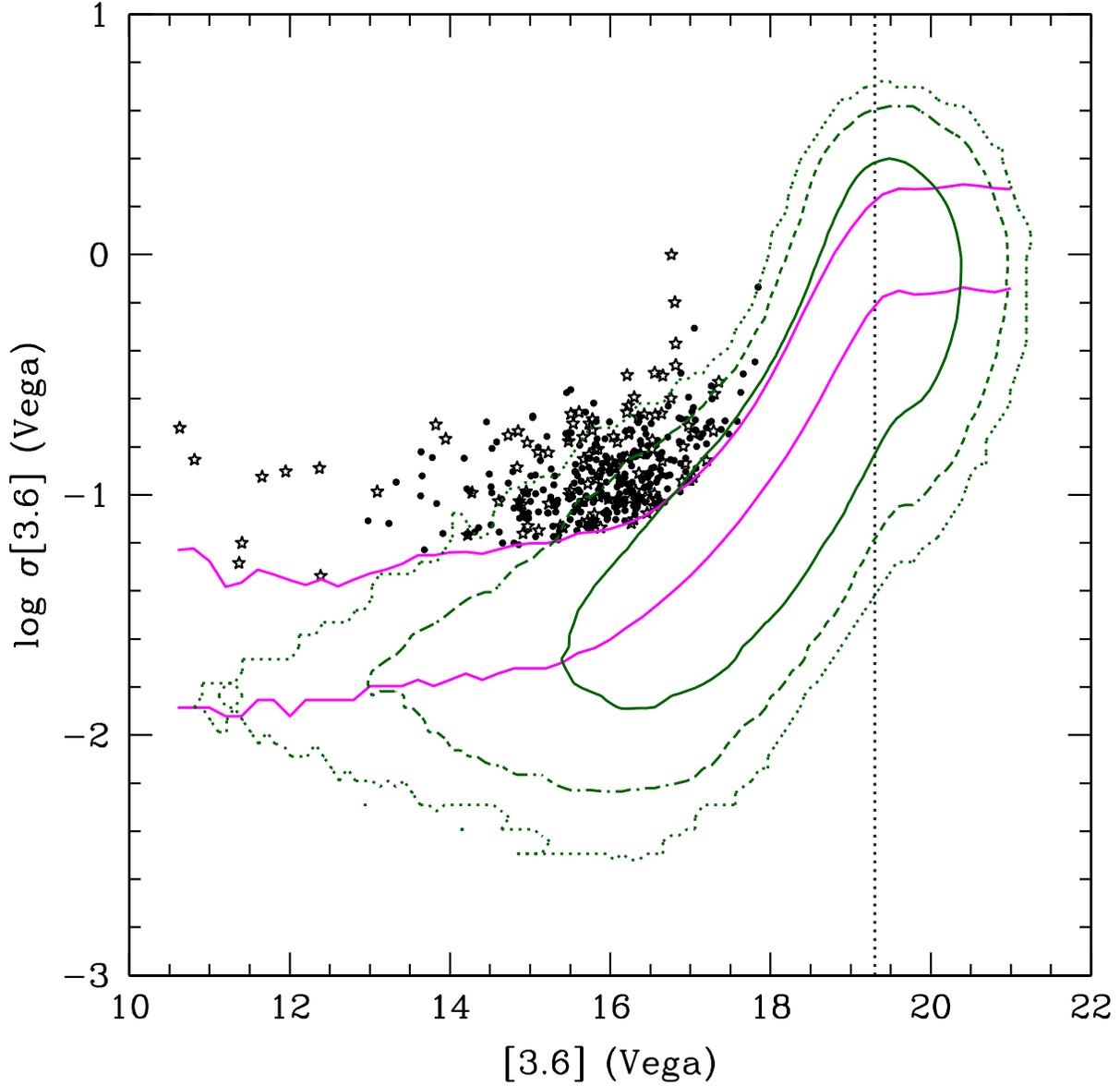}
\caption{The distribution of $\sim440,000$ SDWFS sources in 
magnitude-log($\sigma$) space, where $\sigma$ represents the
variation in photometry over four epochs at 3.6\,\micron.  The sources 
were binned at 0.1\,mag intervals, and within these bins, both the 
median photometric variation ($\bar{\sigma}$) and the dispersion with respect 
to that median ($\hat{\sigma}$) were computed.
The green contours show the magnitude-log($\sigma$) distribution of SDWFS 
sources when binned in 0.1\,mag intervals and 0.05 log($\sigma$) bins, 
indicating 2, 10, and 100 sources per bin.
The median variation $\bar{\sigma}$ and the 1.5\,$\hat{\sigma}$ level of significance 
above $\bar{\sigma}$ are indicated by the lower and upper magenta lines, respectively.
A total of 379 objects satisfy the variability criteria described in the text, i.e.,
$\sigma > 1.5 \hat{\sigma} + \bar{\sigma}$.
Sources that lie within (outside) the AGN wedge are indicated with 
filled circles (open stars).
The vertical dotted line at [3.6]=19.3\,mag marks the limiting magnitude of 
the variability study.  
\label{fig:variables}}
\end{figure*}

\begin{figure*}
\epsscale{1.15}
\plottwo{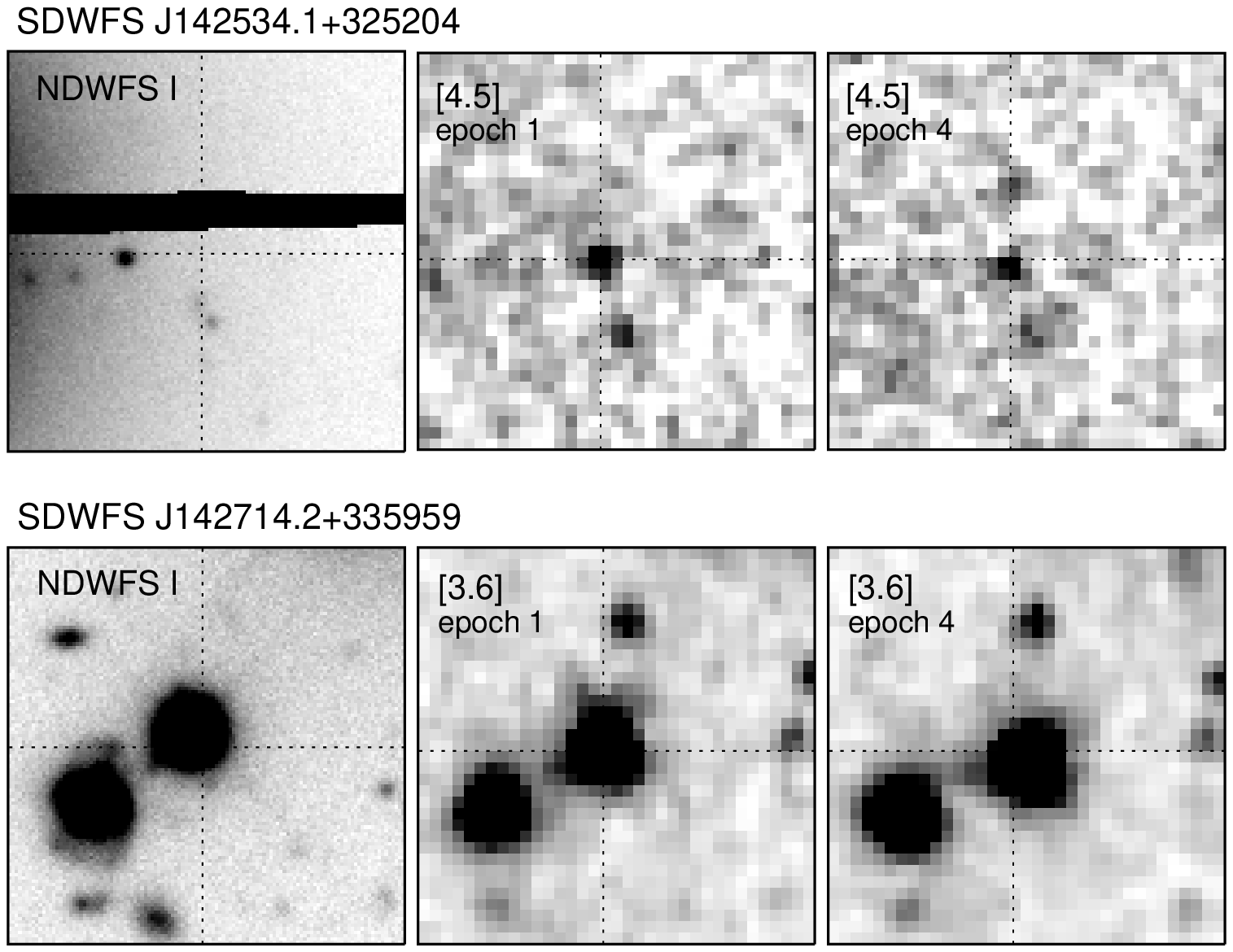}{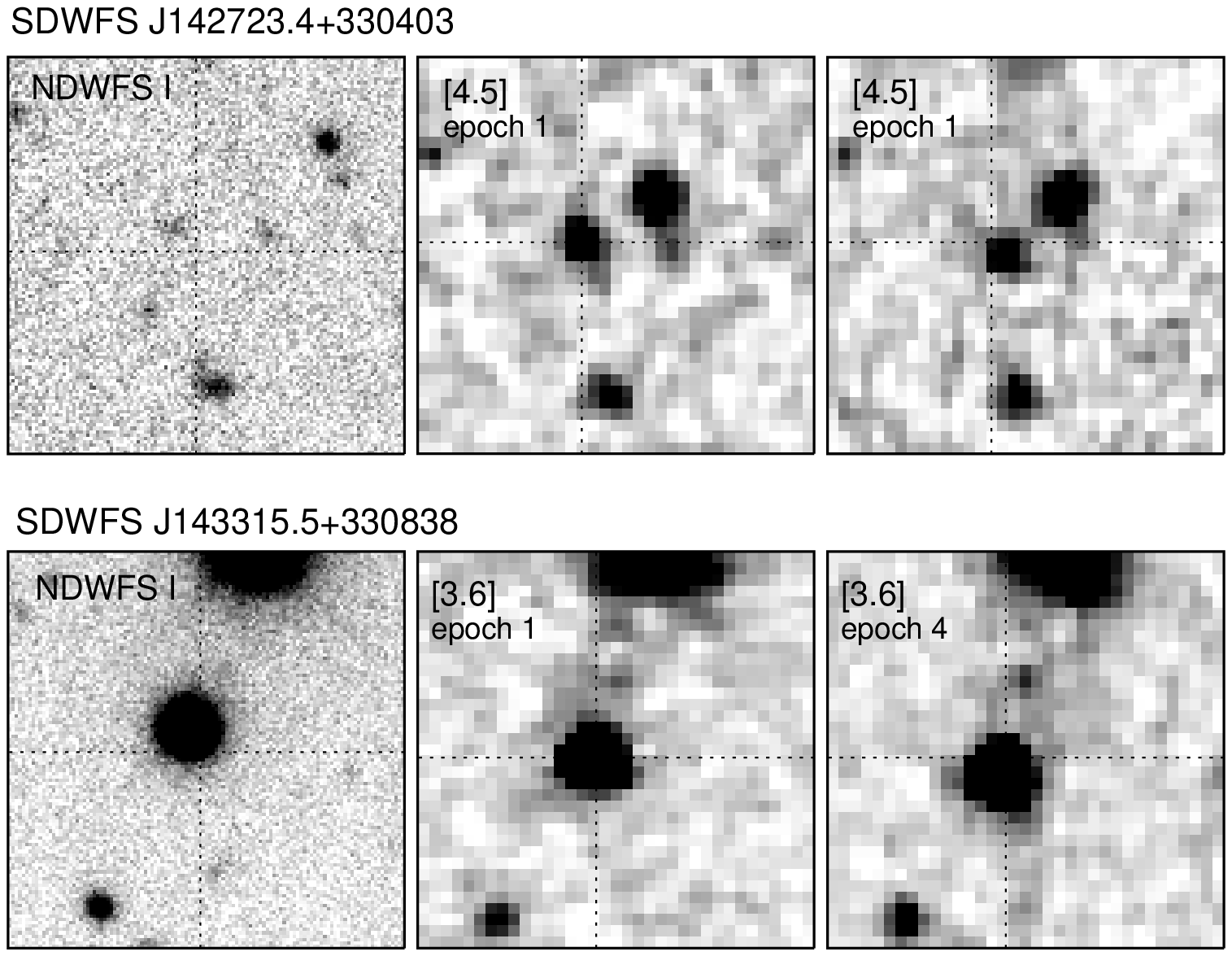}
\caption{Images of the four sources with the highest measured proper
motions across the four-epoch SDWFS campaign.  The left image of
each triple shows the NDWFS $I$-band image (circa 2000), 
the center image shows the first-epoch SDWFS image 
(2004 January), and the right image shows
the fourth-epoch SDWFS image (2008 March).  Images are all 30\arcsec\
on a side, with North up and East to the left.  Cross hairs are
centered on the 2004 January position.  Two of the sources have
colors consistent with mid-T brown dwarfs.  
The two brightest sources were identified as high
proper motion sources at both 3.6 and 4.5\,$\mu$m.  
The two fainter, mid-T candidates
were only identified as high proper motion candidates in the 4.5~$\mu$m mosaics, 
in which they have higher signal-to-noise ratio.
\label{fig:propermotion}}
\end{figure*}

\begin{figure}
\plotone{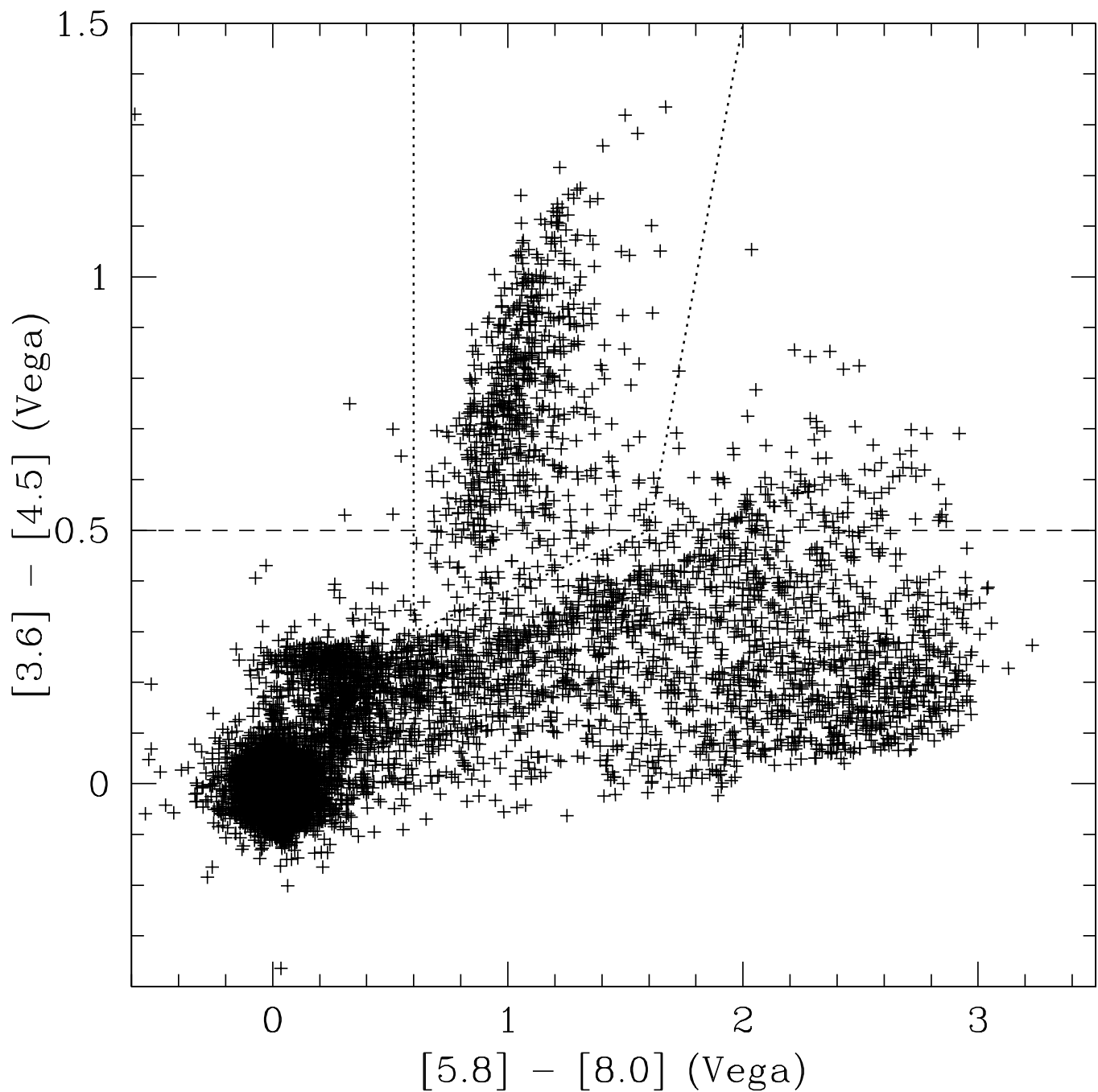}
\caption{The colors of SDWFS sources that lie above the 
expected 5$\sigma$ detection limits of the planned {\sl WISE} all-sky survey.  
The majority of sources redder than 
${\rm [3.6]} - {\rm [4.5]} = 0.5$ have the colors of AGN.
\label{fig:WISEdepth}}
\end{figure}

\rotate{
\begin{deluxetable}{lcccccccccccl}
\tabletypesize{\scriptsize}
\tablewidth{0pt}
\tablecaption{High Proper Motion Sources.}
\tablehead{
\colhead{Target} & \colhead{$\langle \mu_{\rm R.A.} \rangle$} & \colhead{$\langle \mu_{\rm Dec.} \rangle$} &
\colhead{$B_W$} & \colhead{$R$} & \colhead{$I$} &
\colhead{$J$} & \colhead{$K_s$} &
\colhead{[3.6]} & \colhead{[4.5]} & \colhead{[5.8]} & \colhead{[8.0]} &
\colhead{Notes}}
\startdata
J142534.1+325204\tablenotemark{a} 
                 & $-$0\farcs15 & $-$0\farcs54 &\nodata&\nodata&\nodata& \nodata & \nodata & 17.83 & 17.37 &   16.65 &$>15.85$ & $\approx$ T3 brown dwarf \\
J142714.2+335959\tablenotemark{b}
                 & $-$0\farcs31 & $-$0\farcs21 & 19.44 &\nodata&\nodata&   14.89 &   14.19 & 13.81 & 13.75 &   13.59 &   13.61 & $\approx$ M1 \\
J142723.4+330403 & $-$0\farcs30 & $-$0\farcs17 & 25.48 & 23.61 & 23.82 & \nodata & \nodata & 18.21 & 16.96 &$>16.66$ &$>15.85$ & $\approx$ T7 brown dwarf \\
J143315.5+330838 & $-$0\farcs05 & $-$0\farcs35 & 23.56 & 19.74 & 17.77 &   16.11 &   15.41 & 14.98 & 15.04 &   14.85 &   14.90 &           M5 --- West \etal (2008) \\
\enddata
\tablenotetext{a}{Photometry for this source is heavily affected by
the bleed trail from a nearby bright star, but the source appears
undetected in the NDWFS imaging.}
\tablenotetext{b}{J142714.2+335959 is saturated in NDWFS $R$ and $I$.  
It has  $ugriz= 21.24, 18.61, 17.09, 16.35$, and 15.96\,AB mag according
to SDSS DR6.}

\tablecomments{Astrometry, used for the nomenclature, is derived
from the first-epoch 3.6~\micron\ images (obtained during UT 2004
January 10 $-$ 14).  Proper motions are per year, derived by averaging
the epoch~1 to epoch~3 and the epoch~1 to epoch~4 offsets; all four
objects are moving in a SW direction.  All tabulated magnitudes 
are Vega-relative total magnitudes.  Optical photometry is from the
NDWFS ($B_{\rm W}RI$; Jannuzi et al., in preparation).  Near-infrared
photometry is from FLAMEX ($JK_s$; Elston et al. 2006).  Mid-infrared
photometry is from SDWFS.  Non-detections are the average 5$\sigma$
limits for the relevant bands across the entire field.  
Near-IR ellipses indicate that
FLAMEX has not observed these sources.  Assigned T classes 
are based on measured [3.6]-[4.5] colors in accordance with Patten \etal (2006).}

\label{tab.propermotion}
\end{deluxetable}
\normalsize
}

\rotate{
\begin{deluxetable}{cccccccc}
\tabletypesize{\scriptsize}
\tablecaption{SDWFS Four-Epoch, 3.6\,\micron-Selected Catalog\label{tbl:format1}}
\tablehead{
\colhead{RA, Dec (J2000)} & 
\colhead{$m_{4,i}$\tablenotemark{a}} &
\colhead{$\sigma_{4,i}$\tablenotemark{b}} &
\colhead{$m_{6,i}$\tablenotemark{a}} &
\colhead{$\sigma_{6,i}$} &
\colhead{MAG\_AUTO$_i$} &
\colhead{$\sigma_{{\rm MAG\_AUTO},i}$} &
\colhead{Flags\tablenotemark{c}} 
}
\startdata
\tabletypesize{\scriptsize}
216.8795190 +32.1472610 & 17.37 17.10 17.09 15.62 &   0.03  0.04  0.32  0.19 & 17.40 17.26 99.00 15.73 &  0.03  0.07 99.00  0.24 & 17.47 17.58 99.00 16.61 &  0.04  0.12 99.00  0.67 &  0 \\
216.8787756 +32.1490457 & 19.07 99.00 99.00 99.00 &   0.12 99.00 99.00 99.00 & 19.07 99.00 99.00 99.00 &  0.15 99.00 99.00 99.00 & 19.14 99.00 17.71 16.12 &  0.20 99.00  1.07  0.36 &  0 \\
216.8779087 +32.1508507 & 19.00 19.10 17.05 99.00 &   0.11  0.23  0.51 99.00 & 19.07 20.04 16.61 99.00 &  0.15  1.19  0.26 99.00 & 19.23 99.00 16.89 99.00 &  0.21 99.00  0.38 99.00 &  0 \\
\enddata
\tablecomments{The SDWFS full-coadd catalog of sources selected using detection 
within the total
3.6\,\micron\ mosaic.  Within each of the sets of four columns (e.g., $m_{4,i}$) 
the tabulated values correspond to 3.6, 4.5, 5.8, and 8.0\,\micron, respectively.
Entries equal to 99 indicate the source lies below the 5$\sigma$ detection limit.
This printed version is only a sample.  The complete tables are available in the electronic
edition of the Astrophysical Journal.  
All 20 SDWFS catalogs are also available from the {\it Spitzer} Science Center as version
DR1.1; they are version 3.4 according to the internal SDWFS team documentation.  
}
\tablenotetext{a}{Aperture-corrected to 12\arcsec\ radius total Vega magnitudes.}
\tablenotetext{b}{Tabulated errors are measurement errors only and do not account 
for any systematic errors.}
\tablenotetext{c}{SExtractor flag output.}
\end{deluxetable}
}

\clearpage

\begin{deluxetable}{cccccccc}
\tabletypesize{\scriptsize}
\tablecaption{SDWFS Four-Epoch, 4.5\,\micron-Selected Catalog\label{tbl:format2}}
\tablewidth{0pt}
\tablehead{
\colhead{RA, Dec (J2000)} &
\colhead{$m_{4,i}$} &
\colhead{$\sigma_{4,i}$} &
\colhead{$m_{6,i}$} &
\colhead{$\sigma_{6,i}$} &
\colhead{MAG\_AUTO$_i$} &
\colhead{$\sigma_{{\rm MAG\_AUTO},i}$} &
\colhead{Flags}
}
\startdata
217.5665371 +32.1172020 & 99.00 18.39 99.00 16.36 &  99.00  0.13 99.00  0.29 & 99.00 18.53 99.00 16.74 & 99.00  0.17 99.00  0.54 & 99.00 18.73 99.00 17.11 & 99.00  0.13 99.00  0.37 & 24 \\
216.1924704 +32.1071896 & 99.00 17.30 99.00 99.00 &  99.00  0.05 99.00 99.00 & 99.00 17.25 99.00 17.04 & 99.00  0.06 99.00  0.82 & 99.00 17.35 99.00 99.00 & 99.00  0.06 99.00 99.00 & 27 \\
216.2097384 +32.1080726 & 99.00 18.13 99.00 17.66 &  99.00  0.13 99.00  2.01 & 99.00 18.07 99.00 99.00 & 99.00  0.14 99.00 99.00 & 99.00 18.35 99.00 99.00 & 99.00  0.12 99.00 99.00 & 16 \\
\enddata
\tablecomments{As Table~\ref{tbl:format1}, but implementing selection within the 
full, four-epoch 4.5\,\micron\ SDWFS mosaic.
}
\end{deluxetable}

\begin{deluxetable}{cccccccc}
\tabletypesize{\scriptsize}
\tablecaption{SDWFS Four-Epoch, 5.8\,\micron-Selected Catalog\label{tbl:format3}}
\tablewidth{0pt}
\tablehead{
\colhead{RA, Dec (J2000)} &
\colhead{$m_{4,i}$} &
\colhead{$\sigma_{4,i}$} &
\colhead{$m_{6,i}$} &
\colhead{$\sigma_{6,i}$} &
\colhead{MAG\_AUTO$_i$} &
\colhead{$\sigma_{{\rm MAG\_AUTO},i}$} &
\colhead{Flags} 
}
\startdata
216.8846970 +32.1580491 & 17.19 17.00 16.32 16.12 &   0.02  0.04  0.19  0.33 & 17.13 16.94 16.17 16.16 &  0.03  0.05  0.17  0.37 & 17.11 17.28 16.34 99.00 &  0.04  0.11  0.32 99.00 &  0 \\
216.8780169 +32.1638692 & 18.39 17.82 16.29 16.11 &   0.07  0.09  0.18  0.33 & 18.14 17.75 16.05 16.31 &  0.06  0.10  0.16  0.43 & 16.99 17.09 15.43 99.00 &  0.04  0.11  0.16 99.00 &  0 \\
216.8807521 +32.1636509 & 16.40 16.23 15.56 14.86 &   0.01  0.02  0.10  0.11 & 16.17 16.03 15.26 14.79 &  0.01  0.02  0.07  0.10 & 15.58 15.93 14.31 14.97 &  0.02  0.07  0.10  0.37 &  0 \\
\enddata
\tablecomments{As Table~\ref{tbl:format1}, but implementing selection within the 
full, four-epoch 5.8\,\micron\ SDWFS mosaic.
}
\end{deluxetable}

\begin{deluxetable}{cccccccc}
\tabletypesize{\scriptsize}
\tablecaption{SDWFS Four-Epoch, 8.0\,\micron-Selected Catalog\label{tbl:format4}}
\tablewidth{0pt}
\tablehead{
\colhead{RA, Dec (J2000)} &
\colhead{$m_{4,i}$} &
\colhead{$\sigma_{4,i}$} &
\colhead{$m_{6,i}$} &
\colhead{$\sigma_{6,i}$} &
\colhead{MAG\_AUTO$_i$} &
\colhead{$\sigma_{{\rm MAG\_AUTO},i}$} &
\colhead{Flags} 
}
\startdata
216.1899512 +32.1081285 & 99.00 14.43 99.00 14.24 &  99.00  0.00 99.00  0.06 & 99.00 14.42 99.00 14.19 & 99.00  0.00 99.00  0.06 & 20.74 14.50 99.00 14.33 &  2.55  0.01 99.00  0.08 & 16 \\
217.0852837 +32.1234650 & 99.00 15.44 99.00 15.39 &  99.00  0.01 99.00  0.16 & 99.00 15.38 99.00 15.38 & 99.00  0.01 99.00  0.18 & 99.00 15.46 99.00 15.36 & 99.00  0.02 99.00  0.18 &  0 \\
216.1933866 +32.1156223 & 99.00 16.60 99.00 15.48 &  99.00  0.04 99.00  0.22 & 99.00 16.01 99.00 15.24 & 99.00  0.02 99.00  0.15 & 20.64 15.30 99.00 15.45 &  1.98  0.02 99.00  0.24 &  0 \\
\enddata
\tablecomments{As Table~\ref{tbl:format1}, but implementing selection within the 
full, four-epoch 8.0\,\micron\ SDWFS mosaic.
}
\end{deluxetable}

\clearpage

\begin{deluxetable}{cccccccc}
\tabletypesize{\scriptsize}
\tablecaption{SDWFS First-Epoch, 3.6\,\micron-Selected Catalog\label{tbl:format5}}
\tablewidth{0pt}
\tablehead{
\colhead{RA, Dec (J2000)} &
\colhead{$m_{4,i}$} &
\colhead{$\sigma_{4,i}$} &
\colhead{$m_{6,i}$} &
\colhead{$\sigma_{6,i}$} &
\colhead{MAG\_AUTO$_i$} &
\colhead{$\sigma_{{\rm MAG\_AUTO},i}$} &
\colhead{Flags}
}
\startdata
218.2410983 +32.3110349 & 18.80 99.00 99.00 99.00 &   0.16 99.00 99.00 99.00 & 18.84 99.00 99.00 99.00 &  0.21 99.00 99.00 99.00 & 19.11 99.00 99.00 99.00 &  0.33 99.00 99.00 99.00 &  2 \\
218.2296437 +32.3150555 & 18.61 99.00 99.00 99.00 &   0.15 99.00 99.00 99.00 & 18.73 99.00 99.00 99.00 &  0.19 99.00 99.00 99.00 & 18.98 99.00 99.00 99.00 &  0.30 99.00 99.00 99.00 &  0 \\
218.2327133 +32.3136986 & 16.93 99.00 99.00 99.00 &   0.03 99.00 99.00 99.00 & 16.85 99.00 16.91 99.00 &  0.03 99.00  0.82 99.00 & 16.93 99.00 99.00 99.00 &  0.06 99.00 99.00 99.00 &  0 \\
\enddata
\tablecomments{Format as Table~\ref{tbl:format1}, but for the first-epoch SDWFS sources, 
incorporating a selection at 3.6\,\micron\ based on the total SDWFS coadd.
}
\end{deluxetable}

\begin{deluxetable}{cccccccc}
\tabletypesize{\scriptsize}
\tablecaption{SDWFS First-Epoch, 4.5\,\micron-Selected Catalog\label{tbl:format6}}
\tablewidth{0pt}
\tablehead{
\colhead{RA, Dec (J2000)} &
\colhead{$m_{4,i}$} &
\colhead{$\sigma_{4,i}$} &
\colhead{$m_{6,i}$} &
\colhead{$\sigma_{6,i}$} &
\colhead{MAG\_AUTO$_i$} &
\colhead{$\sigma_{{\rm MAG\_AUTO},i}$} &
\colhead{Flags}
}
\startdata
218.1755258 +32.2135628 & 99.00 17.71 99.00 15.70 &  99.00  0.15 99.00  0.40 & 99.00 17.73 99.00 15.39 & 99.00  0.19 99.00  0.35 & 99.00 17.79 99.00 15.55 & 99.00  0.19 99.00  0.31 &  0 \\
218.1709414 +32.2140743 & 99.00 18.16 99.00 16.22 &  99.00  0.22 99.00  2.21 & 99.00 18.33 99.00 15.96 & 99.00  0.33 99.00  0.63 & 99.00 18.62 99.00 16.84 & 99.00  0.40 99.00  1.58 &  3 \\
218.1711818 +32.2155026 & 99.00 16.54 99.00 99.00 &  99.00  0.05 99.00 99.00 & 99.00 16.52 99.00 15.52 & 99.00  0.06 99.00  0.39 & 99.00 16.52 99.00 15.50 & 99.00  0.07 99.00  0.39 &  2 \\
\enddata
\tablecomments{Format as Table~\ref{tbl:format1}, but for the first-epoch SDWFS sources,
incorporating a selection at 4.5\,\micron\ based on the total SDWFS coadd.
}
\end{deluxetable}

\begin{deluxetable}{cccccccc}
\tabletypesize{\scriptsize}
\tablecaption{SDWFS First-Epoch, 5.8\,\micron-Selected Catalog\label{tbl:format7}}
\tablewidth{0pt}
\tablehead{
\colhead{RA, Dec (J2000)} &
\colhead{$m_{4,i}$} &
\colhead{$\sigma_{4,i}$} &
\colhead{$m_{6,i}$} &
\colhead{$\sigma_{6,i}$} &
\colhead{MAG\_AUTO$_i$} &
\colhead{$\sigma_{{\rm MAG\_AUTO},i}$} &
\colhead{Flags} 
}
\startdata
218.2483469 +32.3236056 & 99.00 99.00 14.69 99.00 &  99.00 99.00  0.25 99.00 & 99.00 99.00 12.31 99.00 & 99.00 99.00  0.01 99.00 & 99.00 99.00 10.51 99.00 & 99.00 99.00  0.00 99.00 &  3 \\
216.3420939 +32.3147357 & 16.75 16.31 15.65 14.36 &   0.03  0.04  0.18  0.13 & 16.50 16.18 15.83 14.25 &  0.02  0.04  0.26  0.12 & 16.48 16.23 15.32 14.21 &  0.04  0.07  0.24  0.15 &  0 \\
218.2527709 +32.3242584 & 99.00 99.00 12.59 99.00 &  99.00 99.00  0.02 99.00 & 99.00 99.00 12.10 99.00 & 99.00 99.00  0.01 99.00 & 99.00 99.00 11.50 99.00 & 99.00 99.00  0.01 99.00 &  0 \\
\enddata
\tablecomments{Format as Table~\ref{tbl:format1}, but for the first-epoch SDWFS sources,
incorporating a selection at 5.8\,\micron\ based on the total SDWFS coadd.
}
\end{deluxetable}

\begin{deluxetable}{cccccccc}
\tabletypesize{\scriptsize}
\tablecaption{SDWFS First-Epoch, 8.0\,\micron-Selected Catalog\label{tbl:format8}}
\tablewidth{0pt}
\tablehead{
\colhead{RA, Dec (J2000)} &
\colhead{$m_{4,i}$} &
\colhead{$\sigma_{4,i}$} &
\colhead{$m_{6,i}$} &
\colhead{$\sigma_{6,i}$} &
\colhead{MAG\_AUTO$_i$} &
\colhead{$\sigma_{{\rm MAG\_AUTO},i}$} &
\colhead{Flags} 
}
\startdata
216.2963348 +32.2085586 & 99.00 16.96 99.00 14.43 &  99.00  0.07 99.00  0.15 & 99.00 16.75 99.00 14.12 & 99.00  0.07 99.00  0.10 & 99.00 17.09 99.00 13.61 & 99.00  0.25 99.00  0.13 &  0 \\
218.1546740 +32.2255813 & 99.00 16.50 99.00 15.00 &  99.00  0.05 99.00  0.19 & 99.00 16.45 99.00 15.05 & 99.00  0.06 99.00  0.25 & 99.00 16.32 99.00 14.47 & 99.00  0.08 99.00  0.18 &  0 \\
218.1497132 +32.2287087 & 99.00 17.23 99.00 14.78 &  99.00  0.09 99.00  0.17 & 99.00 17.24 99.00 14.74 & 99.00  0.12 99.00  0.18 & 99.00 17.45 99.00 15.19 & 99.00  0.17 99.00  0.27 &  0 \\
\enddata
\tablecomments{Format as Table~\ref{tbl:format1}, but for the first-epoch SDWFS sources,
incorporating a selection at 8.0\,\micron\ based on the total SDWFS coadd.
}
\end{deluxetable}

\begin{deluxetable}{cccccccc}
\tabletypesize{\scriptsize}
\tablecaption{SDWFS Second-Epoch, 3.6\,\micron-Selected Catalog\label{tbl:format9}}
\tablewidth{0pt}
\tablehead{
\colhead{RA, Dec (J2000)} &
\colhead{$m_{4,i}$} &
\colhead{$\sigma_{4,i}$} &
\colhead{$m_{6,i}$} &
\colhead{$\sigma_{6,i}$} &
\colhead{MAG\_AUTO$_i$} &
\colhead{$\sigma_{{\rm MAG\_AUTO},i}$} &
\colhead{Flags}
}
\startdata
216.8795190 +32.1472610 & 17.45 99.00 99.00 99.00 &   0.05 99.00 99.00 99.00 & 17.49 99.00 17.14 99.00 &  0.06 99.00  1.03 99.00 & 17.61 99.00 16.03 99.00 &  0.09 99.00  0.36 99.00 &  0 \\
216.8779087 +32.1508507 & 18.95 99.00 16.34 99.00 &   0.19 99.00  0.57 99.00 & 19.00 99.00 15.57 99.00 &  0.24 99.00  0.19 99.00 & 19.27 99.00 15.32 99.00 &  0.38 99.00  0.16 99.00 &  0 \\
216.8690172 +32.1546013 & 19.11 99.00 16.68 99.00 &   0.25 99.00  0.76 99.00 & 18.81 99.00 15.90 99.00 &  0.20 99.00  0.26 99.00 & 18.23 99.00 14.44 99.00 &  0.25 99.00  0.13 99.00 &  0 \\
\enddata
\tablecomments{Format as Table~\ref{tbl:format1}, but for the second-epoch SDWFS sources,
incorporating a selection at 3.6\,\micron\ based on the total SDWFS coadd.
}
\end{deluxetable}

\begin{deluxetable}{cccccccc}
\tabletypesize{\scriptsize}
\tablecaption{SDWFS Second-Epoch, 4.5\,\micron-Selected Catalog\label{tbl:format10}}
\tablewidth{0pt}
\tablehead{
\colhead{RA, Dec (J2000)} &
\colhead{$m_{4,i}$} &
\colhead{$\sigma_{4,i}$} &
\colhead{$m_{6,i}$} &
\colhead{$\sigma_{6,i}$} &
\colhead{MAG\_AUTO$_i$} &
\colhead{$\sigma_{{\rm MAG\_AUTO},i}$} &
\colhead{Flags}
}
\startdata
216.7849482 +32.1982162 & 17.94 17.64 16.79 99.00 &   0.07  0.12  1.72 99.00 & 17.81 17.64 16.22 16.31 &  0.08  0.16  0.35  0.95 & 17.87 17.72 15.97 16.76 &  0.09  0.19  0.27  2.00 &  0 \\
216.7747301 +32.2006268 & 18.61 17.54 16.12 14.63 &   0.15  0.13  0.30  0.18 & 18.41 17.40 15.92 14.47 &  0.14  0.13  0.26  0.14 & 18.55 17.57 16.12 14.89 &  0.14  0.14  0.26  0.15 &  0 \\
216.7836897 +32.2008446 & 16.56 16.47 16.27 15.30 &   0.02  0.05  0.37  0.29 & 16.45 16.43 15.93 15.14 &  0.02  0.05  0.26  0.26 & 16.51 16.58 15.60 15.30 &  0.03  0.08  0.22  0.31 &  0 \\
\enddata
\tablecomments{Format as Table~\ref{tbl:format1}, but for the second-epoch SDWFS sources,
incorporating a selection at 4.5\,\micron\ based on the total SDWFS coadd.
}
\end{deluxetable}

\begin{deluxetable}{cccccccc}
\tabletypesize{\scriptsize}
\tablecaption{SDWFS Second-Epoch, 5.8\,\micron-Selected Catalog\label{tbl:format11}}
\tablewidth{0pt}
\tablehead{
\colhead{RA, Dec (J2000)} &
\colhead{$m_{4,i}$} &
\colhead{$\sigma_{4,i}$} &
\colhead{$m_{6,i}$} &
\colhead{$\sigma_{6,i}$} &
\colhead{MAG\_AUTO$_i$} &
\colhead{$\sigma_{{\rm MAG\_AUTO},i}$} &
\colhead{Flags} 
}
\startdata
216.8796119 +32.1544672 & 99.00 99.00 15.66 99.00 &  99.00 99.00  0.22 99.00 & 99.00 99.00 15.40 99.00 & 99.00 99.00  0.16 99.00 & 99.00 99.00 15.12 99.00 & 99.00 99.00  0.15 99.00 &  0 \\
216.8846970 +32.1580491 & 17.24 99.00 15.73 99.00 &   0.04 99.00  0.23 99.00 & 17.20 99.00 15.44 99.00 &  0.05 99.00  0.16 99.00 & 17.21 99.00 14.89 99.00 &  0.08 99.00  0.15 99.00 &  0 \\
216.8807521 +32.1636509 & 16.43 99.00 15.15 99.00 &   0.02 99.00  0.14 99.00 & 16.17 99.00 14.77 99.00 &  0.02 99.00  0.09 99.00 & 15.68 99.00 13.34 99.00 &  0.04 99.00  0.08 99.00 &  0 \\
\enddata
\tablecomments{Format as Table~\ref{tbl:format1}, but for the second-epoch SDWFS sources,
incorporating a selection at 5.8\,\micron\ based on the total SDWFS coadd.
}
\end{deluxetable}

\begin{deluxetable}{cccccccc}
\tabletypesize{\scriptsize}
\tablecaption{SDWFS Second-Epoch, 8.0\,\micron-Selected Catalog\label{tbl:format12}}
\tablewidth{0pt}
\tablehead{
\colhead{RA, Dec (J2000)} &
\colhead{$m_{4,i}$} &
\colhead{$\sigma_{4,i}$} &
\colhead{$m_{6,i}$} &
\colhead{$\sigma_{6,i}$} &
\colhead{MAG\_AUTO$_i$} &
\colhead{$\sigma_{{\rm MAG\_AUTO},i}$} &
\colhead{Flags} 
}
\startdata
216.7747441 +32.2004259 & 18.67 17.69 16.06 14.57 &   0.16  0.14  0.28  0.15 & 18.42 17.47 15.92 14.48 &  0.14  0.14  0.26  0.14 & 18.52 17.60 16.16 14.80 &  0.14  0.14  0.27  0.14 &  0 \\
216.7681272 +32.2105936 & 18.14 18.31 99.00 15.10 &   0.10  0.21 99.00  0.18 & 18.07 19.11 17.30 15.71 &  0.10  0.73  1.42  0.46 & 18.34 99.00 16.76 15.90 &  0.16 99.00  0.74  0.58 &  0 \\
216.9966014 +32.2160076 & 18.12 17.64 16.48 13.88 &   0.08  0.15  0.40  0.22 & 18.13 17.55 16.44 12.24 &  0.11  0.15  0.44  0.02 & 18.26 17.77 16.61 12.33 &  0.12  0.18  0.46  0.02 &  0 \\
\enddata
\tablecomments{Format as Table~\ref{tbl:format1}, but for the second-epoch SDWFS sources,
incorporating a selection at 8.0\,\micron\ based on the total SDWFS coadd.
}
\end{deluxetable}

\begin{deluxetable}{cccccccc}
\tabletypesize{\scriptsize}
\tablecaption{SDWFS Third-Epoch, 3.6\,\micron-Selected Catalog\label{tbl:format13}}
\tablewidth{0pt}
\tablehead{
\colhead{RA, Dec (J2000)} & 
\colhead{$m_{4,i}$} & 
\colhead{$\sigma_{4,i}$} &
\colhead{$m_{6,i}$} &
\colhead{$\sigma_{6,i}$} &
\colhead{MAG\_AUTO$_i$} &
\colhead{$\sigma_{{\rm MAG\_AUTO},i}$} &
\colhead{Flags}
}
\startdata
217.2022991 +32.2039361 & 18.48 99.00 99.00 16.38 &   0.15 99.00 99.00  0.84 & 18.06 99.00 99.00 99.00 &  0.10 99.00 99.00 99.00 & 17.92 99.00 99.00 99.00 &  0.10 99.00 99.00 99.00 &  0 \\
217.2056241 +32.2072504 & 18.30 99.00 99.00 99.00 &   0.11 99.00 99.00 99.00 & 18.07 99.00 99.00 16.23 &  0.11 99.00 99.00  0.96 & 18.16 99.00 99.00 16.22 &  0.12 99.00 99.00  0.71 &  1 \\
217.2033917 +32.2074459 & 17.99 18.39 99.00 15.03 &   0.09  0.34 99.00  0.28 & 17.73 18.27 99.00 14.68 &  0.08  0.33 99.00  0.18 & 17.45 99.00 16.78 15.13 &  0.09 99.00  1.27  0.37 &  0 \\
\enddata
\tablecomments{Format as Table~\ref{tbl:format1}, but for the third-epoch SDWFS sources,
incorporating a selection at 3.6\,\micron\ based on the total SDWFS coadd.
}
\end{deluxetable}

\begin{deluxetable}{cccccccc}
\tabletypesize{\scriptsize}
\tablecaption{SDWFS Third-Epoch, 4.5\,\micron-Selected Catalog\label{tbl:format14}}
\tablewidth{0pt}
\tablehead{
\colhead{RA, Dec (J2000)} &
\colhead{$m_{4,i}$} &
\colhead{$\sigma_{4,i}$} &
\colhead{$m_{6,i}$} &
\colhead{$\sigma_{6,i}$} &
\colhead{MAG\_AUTO$_i$} &
\colhead{$\sigma_{{\rm MAG\_AUTO},i}$} &
\colhead{Flags}
}
\startdata
218.0399336 +32.1355066 & 99.00 17.71 99.00 99.00 &  99.00  0.17 99.00 99.00 & 99.00 17.64 99.00 99.00 & 99.00  0.18 99.00 99.00 & 99.00 17.71 99.00 99.00 & 99.00  0.20 99.00 99.00 &  0 \\
217.0086076 +32.1368397 & 99.00 16.70 99.00 16.17 &  99.00  0.08 99.00  2.30 & 99.00 16.34 99.00 15.54 & 99.00  0.05 99.00  0.42 & 99.00 16.40 99.00 15.13 & 99.00  0.09 99.00  0.36 &  0 \\
218.0409377 +32.1412184 & 99.00 17.85 99.00 99.00 &  99.00  0.19 99.00 99.00 & 99.00 17.84 99.00 99.00 & 99.00  0.22 99.00 99.00 & 99.00 17.98 99.00 99.00 & 99.00  0.23 99.00 99.00 &  0 \\
\enddata
\tablecomments{Format as Table~\ref{tbl:format1}, but for the third-epoch SDWFS sources,
incorporating a selection at 4.5\,\micron\ based on the total SDWFS coadd.
}
\end{deluxetable}

\begin{deluxetable}{cccccccc}
\tabletypesize{\scriptsize}
\tablecaption{SDWFS Third-Epoch, 5.8\,\micron-Selected Catalog\label{tbl:format15}}
\tablewidth{0pt}
\tablehead{
\colhead{RA, Dec (J2000)} &
\colhead{$m_{4,i}$} &
\colhead{$\sigma_{4,i}$} &
\colhead{$m_{6,i}$} &
\colhead{$\sigma_{6,i}$} &
\colhead{MAG\_AUTO$_i$} &
\colhead{$\sigma_{{\rm MAG\_AUTO},i}$} &
\colhead{Flags} 
}
\startdata
217.1870834 +32.2105374 & 99.00 17.63 15.74 16.34 &  99.00  0.17  0.24  1.19 & 99.00 17.45 15.49 16.31 & 99.00  0.15  0.18  1.08 & 17.99 17.54 12.52 14.43 &  0.23  0.39  0.03  0.28 &  0 \\
216.9497752 +32.2168119 & 15.91 15.86 15.59 15.50 &   0.01  0.03  0.20  0.40 & 15.65 15.64 15.50 15.18 &  0.01  0.03  0.19  0.29 & 15.41 15.43 16.21 14.50 &  0.02  0.06  1.27  0.35 &  3 \\
217.1808066 +32.2194348 & 15.45 15.73 15.72 15.14 &   0.01  0.02  0.25  0.26 & 15.41 15.71 15.42 15.13 &  0.01  0.03  0.17  0.28 & 14.98 16.72 15.03 13.53 &  0.02  0.26  0.38  0.17 &  0 \\
\enddata
\tablecomments{Format as Table~\ref{tbl:format1}, but for the third-epoch SDWFS sources,
incorporating a selection at 5.8\,\micron\ based on the total SDWFS coadd.
}
\end{deluxetable}

\begin{deluxetable}{cccccccc}
\tabletypesize{\scriptsize}
\tablecaption{SDWFS Third-Epoch, 8.0\,\micron-Selected Catalog\label{tbl:format16}}
\tablewidth{0pt}
\tablehead{
\colhead{RA, Dec (J2000)} &
\colhead{$m_{4,i}$} &
\colhead{$\sigma_{4,i}$} &
\colhead{$m_{6,i}$} &
\colhead{$\sigma_{6,i}$} &
\colhead{MAG\_AUTO$_i$} &
\colhead{$\sigma_{{\rm MAG\_AUTO},i}$} &
\colhead{Flags} 
}
\startdata
218.0345373 +32.1436154 & 99.00 16.06 99.00 14.91 &  99.00  0.04 99.00  0.21 & 99.00 16.07 99.00 14.88 & 99.00  0.04 99.00  0.22 & 99.00 16.24 99.00 15.06 & 99.00  0.06 99.00  0.24 &  0 \\
216.6480548 +32.1409453 & 99.00 99.00 99.00 13.04 &  99.00 99.00 99.00  0.04 & 99.00 99.00 99.00 13.12 & 99.00 99.00 99.00  0.04 & 99.00 99.00 99.00 13.50 & 99.00 99.00 99.00  0.07 &  0 \\
218.0494313 +32.1458605 & 99.00 15.71 99.00 13.33 &  99.00  0.03 99.00  0.05 & 99.00 15.64 99.00 13.27 & 99.00  0.03 99.00  0.05 & 99.00 15.72 99.00 13.40 & 99.00  0.04 99.00  0.07 &  0 \\
\enddata
\tablecomments{Format as Table~\ref{tbl:format1}, but for the third-epoch SDWFS sources,
incorporating a selection at 8.0\,\micron\ based on the total SDWFS coadd.
}
\end{deluxetable}

\begin{deluxetable}{cccccccc}
\tabletypesize{\scriptsize}
\tablecaption{SDWFS Fourth-Epoch, 3.6\,\micron-Selected Catalog\label{tbl:format17}}
\tablewidth{0pt}
\tablehead{
\colhead{RA, Dec (J2000)} &
\colhead{$m_{4,i}$} &
\colhead{$\sigma_{4,i}$} &
\colhead{$m_{6,i}$} &
\colhead{$\sigma_{6,i}$} &
\colhead{MAG\_AUTO$_i$} &
\colhead{$\sigma_{{\rm MAG\_AUTO},i}$} &
\colhead{Flags}
}
\startdata
216.3064840 +32.1872639 & 18.25 17.71 16.13 16.45 &   0.11  0.16  0.29  0.85 & 18.07 17.49 16.11 16.34 &  0.10  0.14  0.30  0.98 & 18.20 17.61 16.40 16.54 &  0.10  0.14  0.32  0.78 &  3 \\
216.7761606 +32.1925195 & 18.59 19.23 16.54 99.00 &   0.14  0.71  0.44 99.00 & 18.41 19.05 16.20 99.00 &  0.14  0.69  0.33 99.00 & 18.11 19.37 16.42 99.00 &  0.13  2.55  0.50 99.00 &  2 \\
216.7971180 +32.1938184 & 18.46 19.03 99.00 16.21 &   0.14  0.49 99.00  2.01 & 18.04 18.84 99.00 15.11 &  0.10  0.54 99.00  0.25 & 17.50 18.72 99.00 14.42 &  0.08  0.69 99.00  0.15 &  2 \\
\enddata
\tablecomments{Format as Table~\ref{tbl:format1}, but for the fourth-epoch SDWFS sources,
incorporating a selection at 3.6\,\micron\ based on the total SDWFS coadd.
}
\end{deluxetable}

\begin{deluxetable}{cccccccc}
\tabletypesize{\scriptsize}
\tablecaption{SDWFS Fourth-Epoch, 4.5\,\micron-Selected Catalog\label{tbl:format18}}
\tablewidth{0pt}
\tablehead{
\colhead{RA, Dec (J2000)} &
\colhead{$m_{4,i}$} &
\colhead{$\sigma_{4,i}$} &
\colhead{$m_{6,i}$} &
\colhead{$\sigma_{6,i}$} &
\colhead{MAG\_AUTO$_i$} &
\colhead{$\sigma_{{\rm MAG\_AUTO},i}$} &
\colhead{Flags}
}
\startdata
217.3516233 +32.1541013 & 99.00 17.30 99.00 16.17 &  99.00  0.10 99.00  0.69 & 99.00 17.46 99.00 15.87 & 99.00  0.14 99.00  0.54 & 99.00 17.80 99.00 16.19 & 99.00  0.21 99.00  0.67 &  0 \\
217.8643815 +32.1544045 & 99.00 17.08 99.00 15.86 &  99.00  0.07 99.00  0.98 & 99.00 17.18 99.00 15.85 & 99.00  0.11 99.00  0.53 & 99.00 17.32 99.00 15.59 & 99.00  0.14 99.00  0.38 &  3 \\
217.8627125 +32.1547360 & 99.00 15.71 99.00 14.80 &  99.00  0.02 99.00  0.20 & 99.00 15.56 99.00 14.46 & 99.00  0.02 99.00  0.14 & 99.00 15.44 99.00 14.52 & 99.00  0.03 99.00  0.18 &  3 \\
\enddata
\tablecomments{Format as Table~\ref{tbl:format1}, but for the fourth-epoch SDWFS sources,
incorporating a selection at 4.5\,\micron\ based on the total SDWFS coadd.
}
\end{deluxetable}

\begin{deluxetable}{cccccccc}
\tabletypesize{\scriptsize}
\tablecaption{SDWFS Fourth-Epoch, 5.8\,\micron-Selected Catalog\label{tbl:format19}}
\tablewidth{0pt}
\tablehead{
\colhead{RA, Dec (J2000)} &
\colhead{$m_{4,i}$} &
\colhead{$\sigma_{4,i}$} &
\colhead{$m_{6,i}$} &
\colhead{$\sigma_{6,i}$} &
\colhead{MAG\_AUTO$_i$} &
\colhead{$\sigma_{{\rm MAG\_AUTO},i}$} &
\colhead{Flags} 
}
\startdata
216.3143103 +32.1907240 & 16.56 16.31 15.76 15.65 &   0.02  0.04  0.23  0.41 & 16.40 16.14 15.47 15.50 &  0.02  0.04  0.16  0.37 & 15.93 15.73 15.39 99.00 &  0.04  0.09  0.46 99.00 &  2 \\
216.7530185 +32.1992019 & 14.75 14.96 15.00 14.97 &   0.00  0.01  0.11  0.20 & 14.71 14.91 14.82 15.01 &  0.00  0.01  0.09  0.23 & 14.60 14.72 14.71 15.29 &  0.01  0.02  0.15  0.52 &  0 \\
216.2371374 +32.1955547 & 15.49 15.03 14.66 12.75 &   0.01  0.01  0.08  0.03 & 15.33 14.91 14.46 12.68 &  0.01  0.01  0.06  0.03 & 15.02 14.66 13.83 12.49 &  0.02  0.04  0.13  0.07 &  2 \\
\enddata
\tablecomments{Format as Table~\ref{tbl:format1}, but for the fourth-epoch SDWFS sources,
incorporating a selection at 5.8\,\micron\ based on the total SDWFS coadd.
}
\end{deluxetable}

\begin{deluxetable}{cccccccc}
\tabletypesize{\scriptsize}
\tablecaption{SDWFS Fourth-Epoch, 8.0\,\micron-Selected Catalog\label{tbl:format20}}
\tablewidth{0pt}
\tablehead{
\colhead{RA, Dec (J2000)} &
\colhead{$m_{4,i}$} &
\colhead{$\sigma_{4,i}$} &
\colhead{$m_{6,i}$} &
\colhead{$\sigma_{6,i}$} &
\colhead{MAG\_AUTO$_i$} &
\colhead{$\sigma_{{\rm MAG\_AUTO},i}$} &
\colhead{Flags} 
}
\startdata
217.8626725 +32.1545472 & 99.00 15.76 99.00 14.75 &  99.00  0.03 99.00  0.19 & 99.00 15.60 99.00 14.40 & 99.00  0.03 99.00  0.13 & 99.00 15.76 99.00 14.75 & 99.00  0.02 99.00  0.12 &  0 \\
217.3706421 +32.1574812 & 99.00 15.01 99.00 14.49 &  99.00  0.01 99.00  0.13 & 99.00 15.01 99.00 14.40 & 99.00  0.01 99.00  0.13 & 99.00 15.09 99.00 13.94 & 99.00  0.02 99.00  0.11 &  0 \\
217.3174334 +32.1592254 & 99.00 99.00 99.00 14.86 &  99.00 99.00 99.00  0.19 & 99.00 18.62 99.00 14.82 & 99.00  0.42 99.00  0.19 & 99.00 18.99 99.00 15.32 & 99.00  0.44 99.00  0.18 &  0 \\
\enddata
\tablecomments{Format as Table~\ref{tbl:format1}, but for the fourth-epoch SDWFS sources,
incorporating a selection at 8.0\,\micron\ based on the total SDWFS coadd.
}
\end{deluxetable}

\end{document}